\documentclass[twocolumn,superscriptaddress,prB]{revtex4-1}
\usepackage{graphicx}
\usepackage[hidelinks]{hyperref}
\pdfoutput=1

\begin{document}
\title{Strong 1D localization and highly anisotropic electron-hole masses in heavy-halogen functionalized graphenes}

\author{Lukas Eugen \surname{Marsoner Steinkasserer}}
\email{marsoner@zedat.fu-berlin.de}
\affiliation{Institut f\"{u}r Chemie und Biochemie, Freie Universit\"{a}t Berlin, Takustra\ss e 3, D-14195 Berlin, Germany}

\author{Alessandra Zarantonello}
\affiliation{Institut f\"{u}r Chemie und Biochemie, Freie Universit\"{a}t Berlin, Takustra\ss e 3, D-14195 Berlin, Germany}

\author{Beate Paulus}
\affiliation{Institut f\"{u}r Chemie und Biochemie, Freie Universit\"{a}t Berlin, Takustra\ss e 3, D-14195 Berlin, Germany}

\begin{abstract}While halogenation of graphene presents a fascinating avenue to the construction of a chemically and physically diverse class of systems, their application in photovoltaics has been hindered by often prohibitively large optical gaps. Herein we study the effects of partial bromination and chlorination on the structure and optoelectronic properties of both graphane and fluorographene. We find brominated and chlorinated fluorographene derivatives to be as stable as graphane making them likely to be durable even at elevated temperatures. A detailed investigation of the systems band structure reveals significant 1D localization of the charge carriers as well as strongly electron-hole asymmetric effective masses. Lastly using $G_0W_0$ and BSE, we investigate the optical adsorption spectra of the aforementioned materials whose first adsorption peak is shown to lie close to the optimal peak position for photovoltaic applications ($\approx 1.5$ eV).
\end{abstract}

\date{July 27, 2016}
\maketitle

\section{Introduction}

The need for clean and sustainable energy has become one of the main driving forces of scientific research in the $21.$ century. Amongst the large number of proposed solutions to the energy crisis, solar cell technology has received a great amount of attention promising to replace large parts of fossil-fuel based energy production in the future. For this to be realized though, the availability of cheap and efficient solar cell materials is of great importance. While classical solar cell technology is mainly based on silicon, 2D materials, notably graphene\cite{wang2008transparent, miao2012high, wu2008organic, liu2008organic} and MoS$_2$\cite{komsa2012two, ganatra2014few, tsai2014monolayer, gu2013solution} have recently emerged as possible alternatives, allowing for the construction of ultrathin photovoltaic devices\cite{bernardi2013extraordinary}.

While graphene itself has sparked the interest of the photovoltaic community, its chemical modifications, most prominently its fully hydrogenated form (graphane) and its fully fluorinated form (fluorographene), have not yet found applications in solar cell technology due to their prohibitively large optical gaps \cite{karlicky2013band}. Another problem plaguing all possible applications of 2D materials in solar cells is the presence of strongly bound electron-hole pairs (excitons) created upon optical excitation \cite{choi2015linear, latini2015excitons}. To achieve large photocurrents the electron-hole pairs should be easily separable which makes the application of 2D materials to solar cells challenging.

Herein we present a possible solution to both the aforementioned problems via the introduction of heavy halogen atoms into graphane and fluorographene. While, as we will show, these modifications help to significantly redshift the optical adsorption of the aforementioned materials, the strong asymmetry in their charge-carrier masses could be exploited to overcome the problem of strong exciton binding, allowing for an efficient separation of electron-hole pairs and in turn high quantum yields in future solar cells. We will consider in detail some of the main properties required for a viable solar cell i.e.~its stability, optical adsorption spectrum and exciton binding energy using single particle methods i.e.~DFT as well as many-body methods like $GW$\cite{hedin1965new} and the Bethe-Salpeter equation (BSE)\cite{salpeter1951relativistic}, to accurately account for both electron-electron as well as electron-hole interactions. 

Our study is based on earlier work performed by Karlick\'{y}, Zbo\v{z}il and Otyepka\cite{karlicky2012band} though we extend on the structures proposed by them and take a more in-depth look at the systems structural, electronic as well as optical properties. The systems are based on $1\times 2$ supercells of graphane and fluorographene in which every second row of H/F atoms along the zigzag direction has been substituted by Br/Cl and we considered both symmetric as well as asymmetric substitutions and functionalization on only one of the two faces of the graphane/fluorographene layer.

\section{Computational Details}

Structure optimizations on all systems were performed using the CRYSTAL14 program\cite{dovesi2014crystal14,crystal14man} together with the M06-2X\cite{zhao2008m06} functional using the \mbox{POB-triple-$\zeta$} basis set proposed by Peintinger et al.\cite{peintinger2013consistent}. In the case of Br and Cl, HSE03\cite{krukau2006influence, heyd2006erratum, heyd2004efficient, heyd2003hybrid} band gap calculations on the relaxed structures were done employing the Stuttgart triple-$\zeta$ basis set as modified for use in periodic calculations by Steenbergen et al.\cite{steenbergen2014method}, together with the associated quasirelativistic pseudopotentials\cite{dolg1987energy, martin2001correlation}. For C, F and H, basis sets were constructed according to the procedure described by Usvyat\cite{usvyat2015high}. In all cases the description of the vacuum region was enhanced by adding ghost atoms containing a 1s function with an exponent of 0.06 $a_0^ {-1}$, 1 \AA\ above the position of the halogen atoms.

Using the obtained structures we performed DFT, $G_0W_0$ and BSE\cite{salpeter1951relativistic,yan2012optical,olsen2016simple} calculations using the GPAW\cite{bahn2002object,mortensen2005real,enkovaara2010electronic,yan2011linear,huser2013quasiparticle,rasmussen2015efficient} code to account for the effect of both electron-electron as well as electron-hole interactions on the systems optical properties. Given the cost of the $G_0W_0$ calculations at the dense k-grids needed for well-converged BSE results, BSE calculations following  $G_0W_0$ were performed employing the scissor approximation i.e.~shifting the unoccupied DFT bands by the energy difference between the DFT and $G_0W_0$ gap. To avoid confusion, BSE calculations based on DFT orbitals where the unoccupied states have been shifted to reproduce the $G_0W_0$ gap will be labeled $^{G_0W_0}$ e.g. BSE@GLLB-SC$^{G_0W_0}$. We tested the validity of this approach for two smaller ($1\times 1$) test systems and found results of full BSE@$G_0W_0$ calculations to agree to within 0.05 eV with those obtained by applying the scissor approximation.

Lastly, as will be seen later, the PBE\cite{perdew1996generalized} functional severely underestimates band gaps for the systems considered herein and, as we suspect, for systems containing strongly localized electrons in general. It therefore provides a poor starting-point for  $G_0W_0$ calculations which assume the DFT one-particle wavefunctions to be close to the true quasiparticle wavefunctions. A possible solution consists in the use of screened hybrid functionals (e.g. HSE03/HSE06) as a starting point for $G_0W_0$ and this approach has already been successfully applied to a number of systems in the literature\cite{fuchs2007quasiparticle, karlicky2013band}. Such calculations are though quite costly computationally as compared to GGA calculations and quickly become prohibitively expensive for larger systems. To circumvent these problems we investigated a low-cost alternative to hybrid functional calculations using the GLLB-SC functional\cite{kuisma2010kohn} which provides a computationally efficient approximation to the EXX-OEP, resulting in a better description of the electronic ground state for the case of highly localized systems. 

The GLLB-SC functional can further be used to calculate the quasiparticle band gap of an N-electron system i.e.~the difference of the ionization potential and electron affinity as the sum of the Kohn-Sham gap and the derivative discontinuity\cite{perdew1982density, kuisma2010kohn}. This approach has been shown to give band gaps in excellent agreement with experimental results\cite{castelli2012computational} at a computational cost close to that of GGA. GLLB-SC$+\Delta_{xc}$ gaps would therefore seem the ideal starting point for subsequent $G_0W_0$ calculations. As pointed out by Yan et al.\cite{yan2012optical} though inclusion of the derivative discontinuity in the calculation of the dielectric constant at the RPA level, i.e.~excluding electron-hole interactions, leads to a systematic underestimation of static screening. Yan et al. focused on BSE calculations where they showed BSE based on GLLB-SC$+\Delta_{xc}$ orbitals and eigenvalues performed better when excluding the derivative discontinuity in the calculation of the dielectric constant. Still the same is not necessarily true for BSE@$G_0W_0$ using GLLB-SC as a starting point. While an underestimation of the dielectric constant at the  $G_0W_0$ level does lead to an increase in the quasiparticle gap, this in turn will result in decreased screening at the BSE level, increasing the exciton binding energy and thereby redshifting the position of the first excitation peak\cite{choi2015linear, olsen2016simple}. 

Given this uncertainty with regards to the best computational method we have opted to provide results both including and excluding the derivative discontinuity (i.e.~$G_0W_0$@GLLBSC and  $G_0W_0$@GLLB-SC$+\Delta_{xc}$). As we will see later, while results are obviously different, the particular choice of computational method does not influence our overall conclusions.

\section{Results and Discussion}

\subsection{Structural properties and stability}

We mentioned in the introduction that, while some of the structures considered in this work were originally proposed by Karlick\'{y} et al. \cite{karlicky2012band}, others have, to the best of our knowledge, never been studied before. We will therefore begin our discussion by briefly laying out the systems as well as discuss their predicted stability compared to better-known graphene halides.

\begin{figure}
\centering
  \includegraphics[width=0.48\textwidth,angle=0]{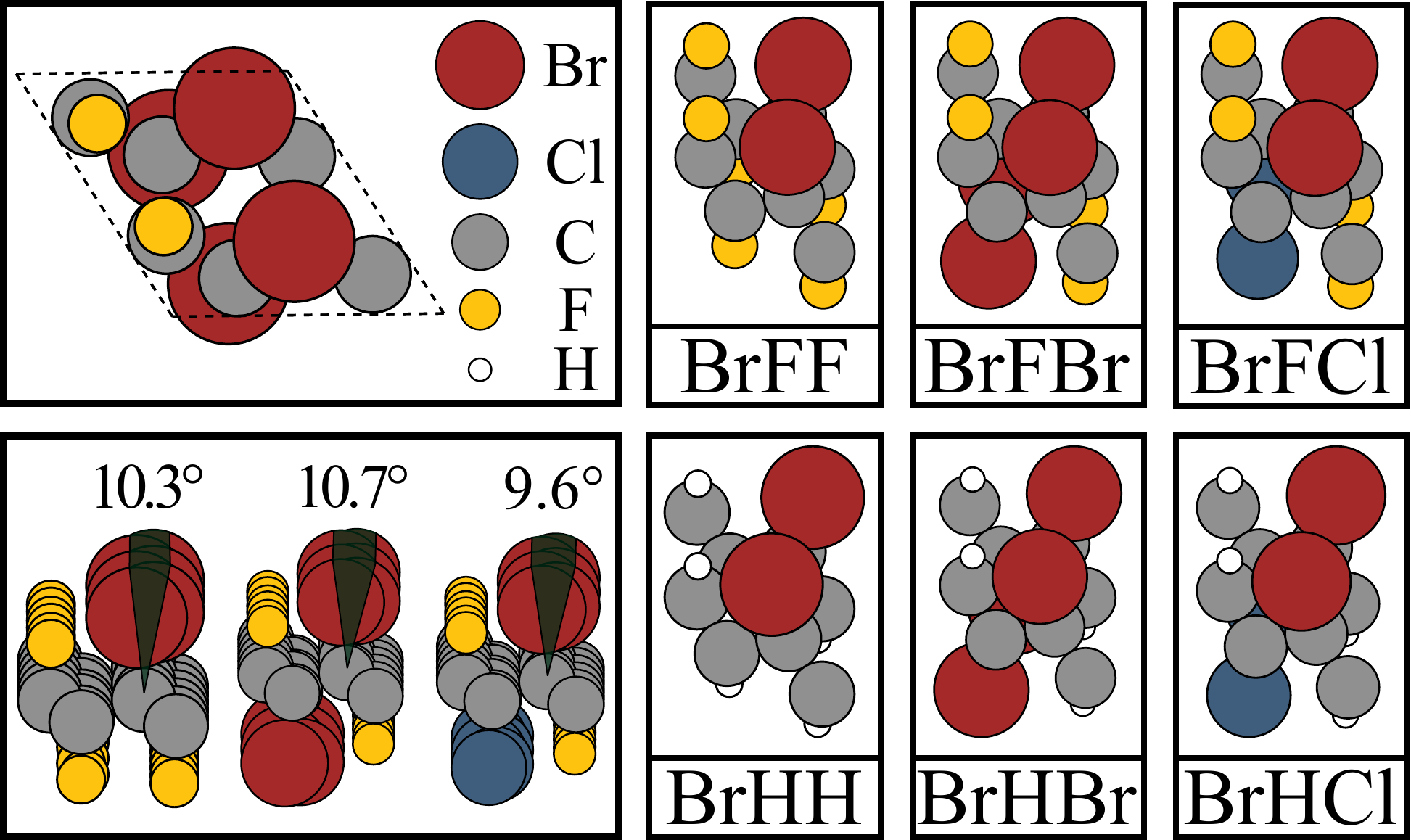}
  \caption{Summary of all systems considered with the corresponding designation which will be used throughout this work. A top-view of the 2x1 supercell construction as well as a site view are also shown. In the latter the Br-buckling as well as the resulting Br-C-C-Br dihedral angle have also been indicated.}
\label{m062x_structures}
\end{figure}

Figure \ref{m062x_structures} shows a schematic representation of all the systems considered in this work as structurally optimized using the M06-2X functional. Comparative structure-relaxations using the PBE functional resulted in only slight differences (see Appendix \ref{supporting}). While the initial study by Karlick\'{y} et al. considered only single unit cells for the pure Br systems (BrFBr and BrHBr) our investigation on $1\times 2$ supercells showed the systems to undergo significant buckling of the bromine atoms. This deformation is indicated for the case of BrFF, BrFBr and BrFCl in figure \ref{m062x_structures}. It results in a notable increase in the band gap, e.g.~ while the BrFBr gap is equal to $\approx0.85$ eV at the GLLB-SC$+\Delta_{xc}$ level if no buckling of the bromines is allowed, the gap increases to 1.52 eV after relaxation in the supercell. This increase is likely caused by the dealignment of the bromine atoms which dominate the systems valence band maximum (VBM). It is consequently stronger in BrFBr as compared to BrFCl where buckling causes an increase in the band gap of $\approx0.5$ eV (going from 0.75 to 1.24 eV) as compared to the BrFBr increase of $\approx0.7$ eV. This difference is in line with the increase in the Br-C-C-Br dihedral angle from $9.6^\circ$ to $10.7^\circ$ as shown in figure \ref{m062x_structures} leading us to believe that it is in fact a dealignment between the Br atoms causing the increase in the band gap.

In order to verify the stability of the systems shown in figure \ref{m062x_structures}, we calculated reaction energies starting from graphane (GrH), fluorographene (GrF) as well as chlorographene (GrCl) and bromographene (GrBr) and the hydrogen/halogen molecules i.e.~H$_2$, F$_2$, Cl$_2$ and Br$_2$ as $E_{\rm{stab.}} = (E_{\rm{Prod.}} - E_{\rm{Reac.}})/N_C$, with $N_C$ being the number of carbon atoms in the system, in analogy to the method used by Karlick\'{y} et al.\cite{karlicky2012band}.

\begin{table}[h]
\centering
\begin{tabular}{@{}lcccc@{}} 
\hline

      &  GrH &  GrF &  GrCl &  GrBr \\ \hline

BrFF  & -67  &  105 &  -203	&  -311 \\
BrFBr &  18  &  189 &  -119	&  -227 \\
BrFCl &  -4  &  167 &  -141	&  -249 \\\hline

BrHH  &  43  &  214 &  -93	&  -202 \\
BrHBr &  69  &  240 &  -68	&  -176 \\
BrHCl &  49  &  220 &  -88	&  -196 \\

\hline
\end{tabular}
 \caption{Stability of the compounds considered in this work as calculated at the M06-2X level. All values in kJ/mol normalized to the number of carbon atoms in the unit cell. For comparison the \emph{M06-2X} stabilities of GrF, GrCl and GrBr relative to GrH are -172 kJ/mol, 136 kJ/mol and 245 kJ/mol respectively.}
\label{stabilities}
\end{table}

While only M06-2X results are shown in table \ref{stabilities}, PBE provides very similar numbers and results are given in the Appendix \ref{supporting}. We see that all systems are more stable than GrCl with fluorographene-base system (BrFF, BrFBr and BrFCl) being more stable than their graphane-based counterparts (BrHH, BrHBr and BrHCl). BrFCl and BrFF in particular are predicted to be more stable than even GrH, making them likely to be durable even at elevated temperatures. Even BrFBr, being only slightly less stable than GrH, could reasonably be expected to resist decomposition under such conditions.

\subsection{Electronic properties}

\begin{figure}
\centering
  \includegraphics[width=0.48\textwidth,angle=0]{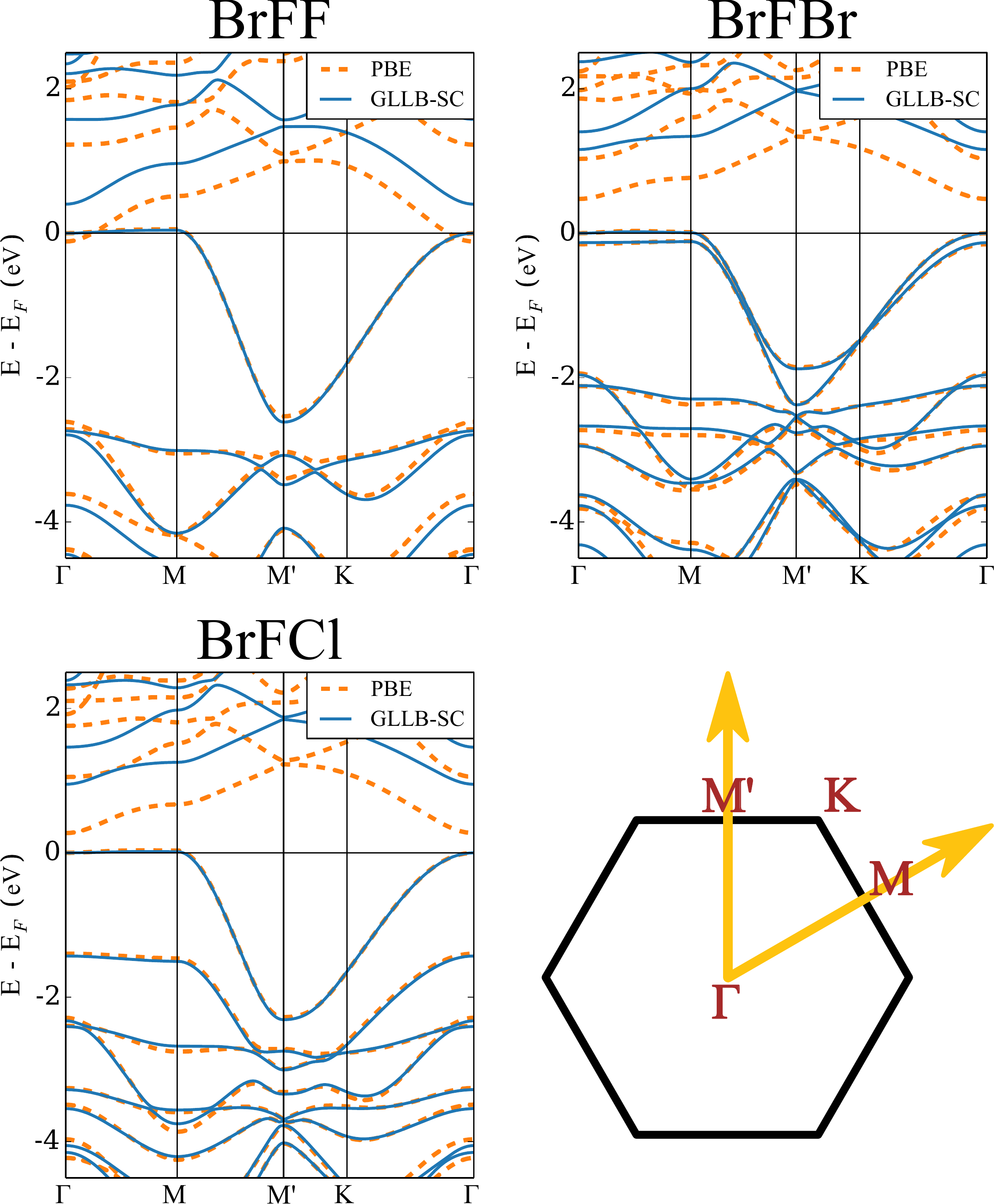}
  \caption{PBE and GLLB-SC band structures for BrFF, BrFBr and BrFCl. The inset on the lower-right further shows a schematic representation of the first Brillouin zone together with the labels for the high-symmetry points used in the band structure plots. Note that we have not included the derivative discontinuity in the figures so as to allow for a direct comparison between the KS-band structures.}
\label{bands_pbe_gllb-sc}
\end{figure}

Having briefly considered the thermodynamic stabilities of the heavy-halogen substituted graphene derivatives we will now turn to their electronic properties. We will focus our detailed analysis on the more stable fluorinated derivatives, beginning with a discussion of their electronic properties by looking at both the PBE as well as the GLLB-SC band structures. The corresponding plots are shown in figure \ref{bands_pbe_gllb-sc} and allow us to make some very interesting observations:

\begin{enumerate}
\item PBE predicts significantly lower band gaps than GLLB-SC in all cases with BrFF being predicted to be conducting at the PBE level while a gap is present in the GLLB-SC results. 
\item All structures show strong band-dispersion along the $\Gamma\rightarrow M'$ and $\Gamma \rightarrow K$ directions while bands close to the Fermi energy ($E_F$) are nearly flat along the $\Gamma\rightarrow M$ direction. 
\item Substitution of Br by Cl leads to an increased splitting of the bands close to the Fermi energy. 
\end{enumerate}

Let us start by considering the first of these observations: Electrons occupying conduction and valence bands close to the Fermi energy are highly localized in all systems as evidenced by the low dispersion of these bands. The failure of PBE to correctly describe the systems band gap is therefore likely attributable to the known failures of GGA-functionals in describing localized systems of electrons\cite{mori2008localization}. GLLB-SC on the other hand, through approximating the OEP-EXX functional, better describes the important on site interactions and yields finite band gaps for all three systems shown in figure \ref{bands_pbe_gllb-sc}.

We now compare band gap values at GLLB-SC and GLLB-SC$+\Delta_{xc}$ level to band gaps calculated using the HSE03 screened hybrid functional within the CRYSTAL14 code. Results are summarized in table \ref{pbe_hse03_gllbsc}. As expected, GLLB-SC is able to correctly reproduce the trends seen in the HSE03 results, predicting all three systems to be semi-conducting. Upon including the derivative discontinuity, the agreement is further improved with band gaps for BrFBr and BrFCl being close to identical in the two methods. The difference is somewhat larger in the case of BrFF, though we are unsure as to what causes this discrepancy. This good agreement between results at the HSE03 and GLLB-SC$+\Delta_{xc}$ level combined with the demonstrated success of HSE03 as a starting point for $G_0W_0$ calculations\cite{fuchs2007quasiparticle, karlicky2013band} makes GLLB-SC$+\Delta_{xc}$ seem to be an excellent starting point for $G_0W_0$ calculations given the low computational cost of GLLB-SC as compared to hybrid-functional calculations. 

\begin{table}[h]
\centering
\begin{tabular}{@{}lccccc@{}} 
\hline
                                 & BrFF & BrFBr & BrFCl \\\hline
HSE03                            & 0.73 & 1.46  & 1.26  \\
GLLB-SC                          & 0.40 & 1.16  & 0.95  \\
GLLB-SC$+\Delta_{xc}$            & 0.52 & 1.54  & 1.25  \\\hline
$G_0W_0$@GLLB-SC                 & 3.14 & 3.52  & 3.47  \\
$G_0W_0$@GLLB-SC$+\Delta_{xc}$   & 3.29 & 3.96  & 3.82  \\
\hline
\end{tabular}
 \caption{Direct fundamental gaps for fluorinated systems using different DFT functionals and methods. While HSE03 calculations are performed using an LCAO basis within the CRYSTAL14 code, all other calculations employ a plane-wave basis and are performed using the GPAW program.}
\label{pbe_hse03_gllbsc}
\end{table}

We stress though that, lacking experimental validation, it is unclear whether or not GLLB-SC$+\Delta_{xc}$ underestimates the macroscopic dielectric constants in our systems as has been shown for both HSE03\cite{fuchs2007quasiparticle} as well as GLLB-SC$+\Delta_{xc}$\cite{yan2012optical} in a series of other materials and if so, how this affects the quality of $G_0W_0$ results. For this reasons we have performed both $G_0W_0$@GLLB-SC as well as $G_0W_0$@GLLB-SC$+\Delta_{xc}$ calculations which might serve as two limiting cases for the true quasiparticle gap. The results of these are also shown in table \ref{pbe_hse03_gllbsc} and, unsurprisingly, they show a further opening of the $G_0W_0$ gap upon inclusion of the derivative discontinuity. 

\begin{figure}
\centering
  \includegraphics[width=0.48\textwidth,angle=0]{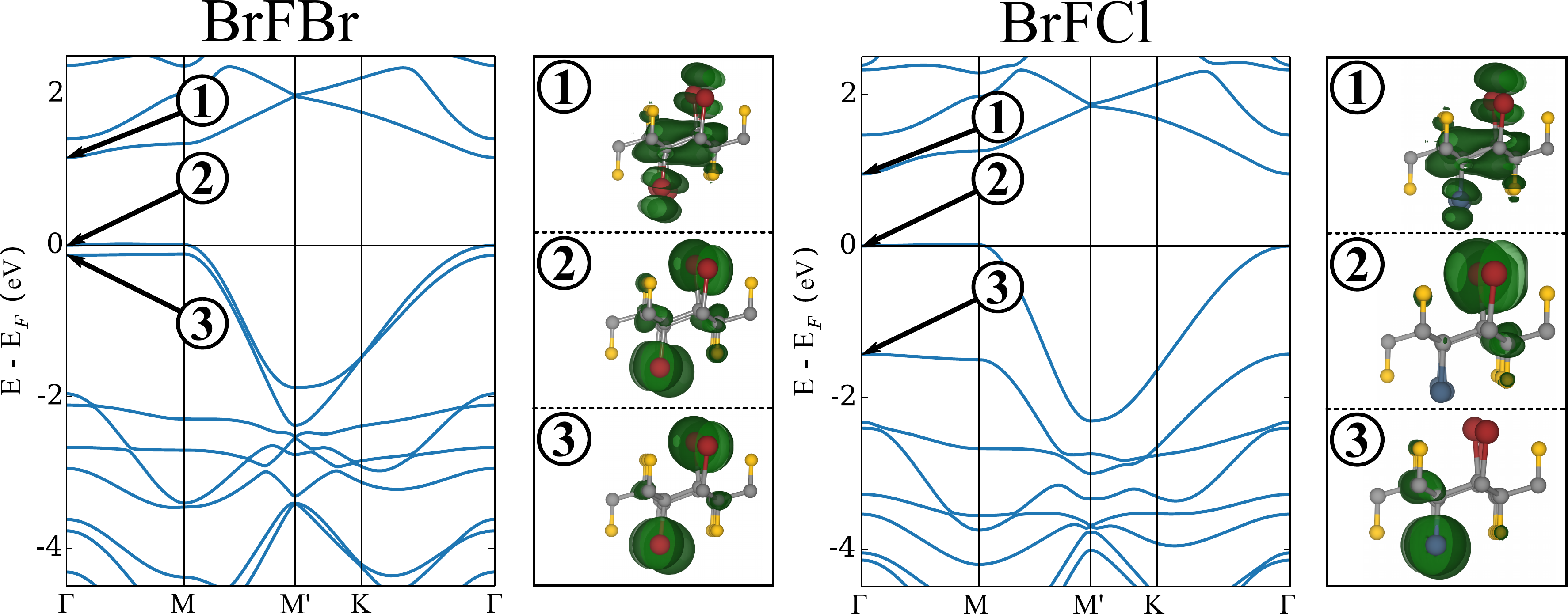}
  \caption{DFT band structures at the GLLB-SC level for the BrFBr and BrFCl system are shown together with charge-density isosurfaces at a level of $8\times10^{-2}$ \AA$^{-3}$.}
\label{local_bands}
\end{figure}

Returning now to the band structures shown in figure \ref{bands_pbe_gllb-sc} it is worth considering the nature of the highly-localized states close to the Fermi energy ($E_F$). To do so in figure \ref{local_bands} we have shown the band structures of BrFBr and BrFCl together with charge-density plots representing the three bands close to $E_F$ at the $\Gamma$-point. In BrFBr both of the closely-spaced occupied bands show strong localization on the chains of bromine atoms along the 2D structure while the unoccupied band forming the conduction band minimum (CBM) is largely delocalized over the entirety of the layer. Upon $\rm{Br}\rightarrow\rm{Cl}$ substitution, while the band forming the CBM is not visibly altered, the two closely-spaced occupied bands lying close to $E_F$ in BrFBr are split significantly. The origin of this split can again be understood from looking at the charge density plots. While the two bands in question are delocalized over the Br atoms lying on both sides of the 2D layer in BrFBr, they become localized on only one site in the case of BrFCl with the band localized on the Cl-side being pushed down in energy with respect to the Br-localized one.

\begin{table}[h]
\centering
\begin{tabular}{@{}lccc@{}} 
\hline
        &                        &       \multicolumn{2}{c}{GLLB-SC}                   \\
\cline{3-4}            

        &                         &  $m_-/m_e$      &   $m_+/m_e$   \\ \hline
BrFF    &  $\Gamma\rightarrow M$  &      0.9        &      51.2     \\
        &  $\Gamma\rightarrow M'$ &      0.6        &       0.3     \\ \hline

BrFBr   &  $\Gamma\rightarrow M$  &      1.6        &      14.0     \\
        &  $\Gamma\rightarrow M'$ &      0.8        &       0.3     \\ \hline

BrFCl   &  $\Gamma\rightarrow M$  &      1.1        &      44.1     \\
        &  $\Gamma\rightarrow M'$ &      0.7        &       0.4     \\ \hline

\end{tabular}
 \caption{Effective masses for electrons ($m_-$) and holes ($m_+$) calculated using GLLB-SC. All values are given as a multiple of the electron rest mass $m_e$. GLLB-SC$+\Delta_{xc}$ results are identical as the derivative discontinuity results only in a rigid shift of the unoccupied bands, leaving their curvature unaffected.}
\label{effective_masses}
\end{table}

Having analyzed the local nature of the bands close to $E_F$ qualitatively we now move to a more quantitative assessment by considering the associated charge carrier effective masses. Here we have obtained electron ($m_-$) and hole ($m_+$) effective masses using the GLLB-SC functional by fitting B-splines to bands at the CBM/VBM along the high-symmetry directions. The effective masses are then obtained by computing the corresponding second derivatives. The results of this analysis are shown in table \ref{effective_masses}. It is worth mentioning at this point that tests on the $1 \times 1$ systems show that the inclusion of electron-electron interactions within $G_0W_0$ does not significantly alter the results obtained using GLLB-SC.

The effective masses now allow us to draw some interesting conclusions regarding the behavior of electrons/holes created upon photoexcitation: electrons are largely unconstrained along both $\Gamma\rightarrow M$ as well as $\Gamma\rightarrow M'$ and so their density will rapidly delocalized over the 2D layer. Holes on the other hand, while having low effective masses along the $\Gamma\rightarrow M'$ direction, are heavily constrained along $\Gamma\rightarrow M$ with effective masses being around two orders of magnitude higher than those along $\Gamma\rightarrow M'$. This strong anisotropy in effective masses will result in rapid delocalization of the hole-density along $\Gamma\rightarrow M'$ (i.e.~the direction \emph{parallel} to the rows of Br atoms) combined with strong localization along $\Gamma\rightarrow M$ (i.e.~the direction \emph{orthogonal} to the rows of Br atoms). This behavior is not only interesting in and of itself but might conceivably be exploited in splitting excitons created within the systems by appropriately varying the external potential in the two spatial directions. For completeness we mention that the band gap in all structures is technically indirect with the VBM lying between $M$ and $\Gamma$ though this does not significantly influence our conclusions given the flatness of the bands along that same direction.

\section*{Optical properties}

\begin{figure}
\centering
  \includegraphics[width=0.48\textwidth,angle=0]{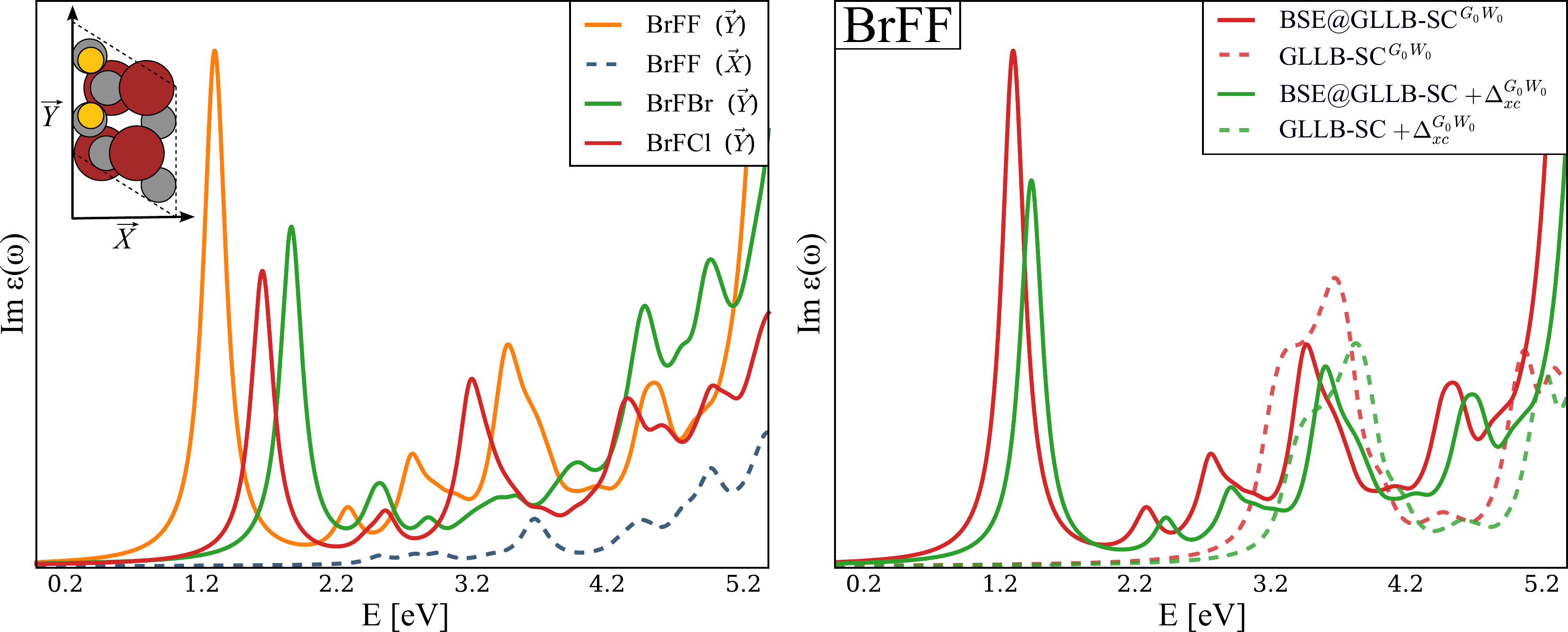}
  \caption{The left-hand figure shows the imaginary part of the frequency dependent dielectric function (Im $\varepsilon(\omega)$) as calculated at the BSE@GLLB-SC$^{G_0W_0}$ level for BrFF, BrFBr and BrFCl. For the case of BrFF spectra obtained for two orthogonal directions of the incident photons are shown. The inset in the left-hand figure shows the atomic structure of BrFF together with the definition of the two directions of incident photons used in the plot. The right-hand figure on the other hand shows the calculated optical spectrum of BrFF at four different levels of theory. In all cases spectra are obtained applying a Lorentzian broadening of 0.1 eV.}
\label{bse_peaks}
\end{figure}

Since we are interested in the optical properties of our systems in particular as they relate to possible applications in photovoltaics, effects due to exciton binding cannot be neglected, especially as they are expected to be of even greater importance in low-dimensional systems such as those considered here\cite{choi2015linear, latini2015excitons} as compared to 3D ones. We therefore performed BSE calculations employing both $G_0W_0$@GLLB-SC as well as  $G_0W_0$@GLLB-SC$+\Delta_{xc}$ as a starting point. Figure \ref{bse_peaks} shows the resulting spectra in the optical limit ($q \rightarrow 0$) at different levels of theory and for different directions of the incoming photons. As expected from the systems band structure shown in figure \ref{bands_pbe_gllb-sc}, the lowest-lying exciton is localized along the $\vec{Y}$-direction while the first peak along $\vec{X}$ lies at $\approx$ 2.53 eV showing very low intensity as compared to the peak along $\vec{Y}$. Comparing the curves shown on the right-hand side of figure \ref{bse_peaks} we can further see that, as expected for a 2D material, electron-hole interactions play a significant role in determining the position of the first optical adsorption peaks with spectra at the $G_0W_0$ level being strongly blue-shifted compared to their BSE counterparts.

\begin{table}
\centering
\begin{tabular}{@{}lccc@{}} 
\hline
                                          &  BrFF  &   BrFBr &   BrFCl   \\ \hline
$G_0W_0$@GLLB-SC                          &  3.14  &   3.52  &   3.47    \\
$G_0W_0$@GLLB-SC$+\Delta_{xc}$            &  3.29  &   3.96  &   3.82    \\\hline
\emph{BSE}@GLLB-SC$^{G_0W_0}$             &  1.31  &   1.88  &   1.67    \\
\emph{BSE}@GLLB-SC$+\Delta_{xc}^{G_0W_0}$ &  1.45  &   2.27  &   1.98    \\
\hline
\end{tabular}
 \caption{Direct $G_0W_0$ band gaps and positions of the lowest-lying adsorption peaks for all fluorographene-based systems shown in figure \ref{m062x_structures}.}
\label{gw_gaps_peaks}
\end{table}

Another interesting observation that can be made from table \ref{gw_gaps_peaks} is the fact that substituting one of the bromine atoms in the unit cell by chlorine and finally fluorine i.e.~moving from BrFBr to BrFCl and BrFF, significantly affects the position of the first excitation peak. While $\rm{Br}\rightarrow\rm{F}$ substitution leaves the exciton binding energy (i.e.~the difference between the fundamental and the optical gap) nearly unaltered at $\approx 1.8$ eV interchanging Br with Cl lowers it by $\approx 0.2$ eV. This change is made all the more interesting by the fact that the change in the $G_0W_0$ gap is $\approx 0.4$ eV upon $\rm{Br}\rightarrow\rm{F}$ substitution while the $G_0W_0$@GLLB-SC gaps of BrFBr and BrFCl only differ by $0.05$ eV. To understand the origin of the effect we return briefly to the discussion of the band structures shown in figure \ref{local_bands}. While pure bromine systems show two almost degenerate bands at the Fermi energy, one of these bands is significantly lowered in energy upon substituting bromine by chlorine which leads to the exciton becoming localized in only one of the two bands. As this band is localized on only one side of the 2D layer in BrFCl as compared to both sides of the layer in BrFBr, the spatial localization of the exciton increases which in turn leads to an increase in the exciton binding energy. Finally, we note that the first adsorption peak for all three systems lies close to the optimal peak position for photovoltaic applications at $\approx 1.5$ eV\cite{shockley1961detailed,miller2012strong}. In particular BrFF, which shows a first adsorption peak at $\approx 1.3$ eV - 1.5 eV (depending on the level of theory) and displays predicted stability higher than that of graphane constitutes a very promising candidate for application in future solar technology.

\section{Conclusions}
Herein we have studied the thermodynamic stability and optoelectronic properties of a series of heavy-halogen substituted graphane and fluorographene derivatives, based on the work originally done by Karlick\'{y}, Zbo\v{z}il and Otyepka.\cite{karlicky2012band}. Graphane-based systems show predicted stabilities lower then graphane, whilst still exceeding that of chlorographane. Fluorographene-based systems on the other hand were shown to display stabilities on par with and even exceeding that of graphane. 

We studied in detail the electronic structure of the aforementioned systems, showing them to display strong 1D-localization of charge carriers with a marked electron-hole asymmetry which we hypothesize might be exploited in separating electron-hole pairs created upon photoexcitation. 

Employing the GLLB-SC functional we were able to obtain band gaps which closely reproduce HSE03 results while keeping computational costs to approximately those of GGA calculations. Given the success of the  $G_0W_0$@HSE03 approach in predicting the band gaps of a number of materials\cite{fuchs2007quasiparticle} we believe it to provide an excellent starting point for $G_0W_0$ calculations, combining low computational requirements with good accuracy. 

Lastly we investigated the optical spectra of the systems including electron-hole interactions on top of  $G_0W_0$ by solving the BSE equation. The first optical adsorption peak for all the systems considered  lies close to the optimal peak position given by the Shockley-Queisser limit i.e.~$\approx 1.5$ eV\cite{shockley1961detailed,miller2012strong}. Especially BrFF, who's first adsorption peak lies at $\approx 1.3 - 1.5$ eV and which shows a predicted stability higher than that of graphane constitutes a very promising candidate for application in future solar technology.

\begin{acknowledgments}
LEMS acknowledges financial support by the Studienstiftung des deutschen Volkes e.V., the Deutsche Forschungsgemeinschaft within the Priority Program (SPP) 1459 (Graphene) as well as the International Max Planck Research School "Complex Surfaces in Material Sciences". The High Performance Computing Network of Northern Germany (HLRN) and computer facilities of the Freie Universit\"{a}t Berlin (ZEDAT) are acknowledged for computer time. The authors are indebted to Kirsten Tr{\o}strup Winther and Filip Anselm Rasmussen (both Copenhagen) for providing the latest development version of the GPAW-$GW$ code and help with its use. We also like to express our gratitude towards Thomas Olsen (Copenhagen) for many fruitful discussions regarding the BSE-calculations. Lastly we thank and Johannes Voss (SUNCAT), Jean Christophe Tremblay, Johannes Budau, Gunter Hermann, Vincent Pohl and Marcel Quennet (all Berlin) for enlightening discussions and suggestions regarding this manuscript. The ASE package \cite{bahn2002object} was used to create images of atomic structures throughout this work, VESTA\cite{momma2011vesta} was used to visualize charge densities, while plots were created using Matplotlib\cite{Hunter_2007}.
\end{acknowledgments}

\bibliography{bibliography}{}

\begin{thebibliography}{47}%
\makeatletter
\providecommand \@ifxundefined [1]{%
 \@ifx{#1\undefined}
}%
\providecommand \@ifnum [1]{%
 \ifnum #1\expandafter \@firstoftwo
 \else \expandafter \@secondoftwo
 \fi
}%
\providecommand \@ifx [1]{%
 \ifx #1\expandafter \@firstoftwo
 \else \expandafter \@secondoftwo
 \fi
}%
\providecommand \natexlab [1]{#1}%
\providecommand \enquote  [1]{``#1''}%
\providecommand \bibnamefont  [1]{#1}%
\providecommand \bibfnamefont [1]{#1}%
\providecommand \citenamefont [1]{#1}%
\providecommand \href@noop [0]{\@secondoftwo}%
\providecommand \href [0]{\begingroup \@sanitize@url \@href}%
\providecommand \@href[1]{\@@startlink{#1}\@@href}%
\providecommand \@@href[1]{\endgroup#1\@@endlink}%
\providecommand \@sanitize@url [0]{\catcode `\\12\catcode `\$12\catcode
  `\&12\catcode `\#12\catcode `\^12\catcode `\_12\catcode `\%12\relax}%
\providecommand \@@startlink[1]{}%
\providecommand \@@endlink[0]{}%
\providecommand \url  [0]{\begingroup\@sanitize@url \@url }%
\providecommand \@url [1]{\endgroup\@href {#1}{\urlprefix }}%
\providecommand \urlprefix  [0]{URL }%
\providecommand \Eprint [0]{\href }%
\providecommand \doibase [0]{http://dx.doi.org/}%
\providecommand \selectlanguage [0]{\@gobble}%
\providecommand \bibinfo  [0]{\@secondoftwo}%
\providecommand \bibfield  [0]{\@secondoftwo}%
\providecommand \translation [1]{[#1]}%
\providecommand \BibitemOpen [0]{}%
\providecommand \bibitemStop [0]{}%
\providecommand \bibitemNoStop [0]{.\EOS\space}%
\providecommand \EOS [0]{\spacefactor3000\relax}%
\providecommand \BibitemShut  [1]{\csname bibitem#1\endcsname}%
\let\auto@bib@innerbib\@empty
\bibitem [{\citenamefont {Wang}\ \emph {et~al.}(2008)\citenamefont {Wang},
  \citenamefont {Zhi},\ and\ \citenamefont {M{\"u}llen}}]{wang2008transparent}%
  \BibitemOpen
  \bibfield  {author} {\bibinfo {author} {\bibfnamefont {X.}~\bibnamefont
  {Wang}}, \bibinfo {author} {\bibfnamefont {L.}~\bibnamefont {Zhi}}, \ and\
  \bibinfo {author} {\bibfnamefont {K.}~\bibnamefont {M{\"u}llen}},\
  }\href@noop {} {\bibfield  {journal} {\bibinfo  {journal} {Nano Lett.}\
  }\textbf {\bibinfo {volume} {8}},\ \bibinfo {pages} {323} (\bibinfo {year}
  {2008})}\BibitemShut {NoStop}%
\bibitem [{\citenamefont {Miao}\ \emph {et~al.}(2012)\citenamefont {Miao},
  \citenamefont {Tongay}, \citenamefont {Petterson}, \citenamefont {Berke},
  \citenamefont {Rinzler}, \citenamefont {Appleton},\ and\ \citenamefont
  {Hebard}}]{miao2012high}%
  \BibitemOpen
  \bibfield  {author} {\bibinfo {author} {\bibfnamefont {X.}~\bibnamefont
  {Miao}}, \bibinfo {author} {\bibfnamefont {S.}~\bibnamefont {Tongay}},
  \bibinfo {author} {\bibfnamefont {M.~K.}\ \bibnamefont {Petterson}}, \bibinfo
  {author} {\bibfnamefont {K.}~\bibnamefont {Berke}}, \bibinfo {author}
  {\bibfnamefont {A.~G.}\ \bibnamefont {Rinzler}}, \bibinfo {author}
  {\bibfnamefont {B.~R.}\ \bibnamefont {Appleton}}, \ and\ \bibinfo {author}
  {\bibfnamefont {A.~F.}\ \bibnamefont {Hebard}},\ }\href@noop {} {\bibfield
  {journal} {\bibinfo  {journal} {Nano Lett.}\ }\textbf {\bibinfo {volume}
  {12}},\ \bibinfo {pages} {2745} (\bibinfo {year} {2012})}\BibitemShut
  {NoStop}%
\bibitem [{\citenamefont {Wu}\ \emph {et~al.}(2008)\citenamefont {Wu},
  \citenamefont {Becerril}, \citenamefont {Bao}, \citenamefont {Liu},
  \citenamefont {Chen},\ and\ \citenamefont {Peumans}}]{wu2008organic}%
  \BibitemOpen
  \bibfield  {author} {\bibinfo {author} {\bibfnamefont {J.}~\bibnamefont
  {Wu}}, \bibinfo {author} {\bibfnamefont {H.~A.}\ \bibnamefont {Becerril}},
  \bibinfo {author} {\bibfnamefont {Z.}~\bibnamefont {Bao}}, \bibinfo {author}
  {\bibfnamefont {Z.}~\bibnamefont {Liu}}, \bibinfo {author} {\bibfnamefont
  {Y.}~\bibnamefont {Chen}}, \ and\ \bibinfo {author} {\bibfnamefont
  {P.}~\bibnamefont {Peumans}},\ }\href@noop {} {\bibfield  {journal} {\bibinfo
   {journal} {Appl. Phys. Lett.}\ }\textbf {\bibinfo {volume} {92}},\ \bibinfo
  {pages} {263302} (\bibinfo {year} {2008})}\BibitemShut {NoStop}%
\bibitem [{\citenamefont {Liu}\ \emph {et~al.}(2008)\citenamefont {Liu},
  \citenamefont {Liu}, \citenamefont {Huang}, \citenamefont {Ma}, \citenamefont
  {Yin}, \citenamefont {Zhang}, \citenamefont {Sun},\ and\ \citenamefont
  {Chen}}]{liu2008organic}%
  \BibitemOpen
  \bibfield  {author} {\bibinfo {author} {\bibfnamefont {Z.}~\bibnamefont
  {Liu}}, \bibinfo {author} {\bibfnamefont {Q.}~\bibnamefont {Liu}}, \bibinfo
  {author} {\bibfnamefont {Y.}~\bibnamefont {Huang}}, \bibinfo {author}
  {\bibfnamefont {Y.}~\bibnamefont {Ma}}, \bibinfo {author} {\bibfnamefont
  {S.}~\bibnamefont {Yin}}, \bibinfo {author} {\bibfnamefont {X.}~\bibnamefont
  {Zhang}}, \bibinfo {author} {\bibfnamefont {W.}~\bibnamefont {Sun}}, \ and\
  \bibinfo {author} {\bibfnamefont {Y.}~\bibnamefont {Chen}},\ }\href@noop {}
  {\bibfield  {journal} {\bibinfo  {journal} {Adv. Mater.}\ }\textbf {\bibinfo
  {volume} {20}},\ \bibinfo {pages} {3924} (\bibinfo {year}
  {2008})}\BibitemShut {NoStop}%
\bibitem [{\citenamefont {Komsa}\ \emph {et~al.}(2012)\citenamefont {Komsa},
  \citenamefont {Kotakoski}, \citenamefont {Kurasch}, \citenamefont {Lehtinen},
  \citenamefont {Kaiser},\ and\ \citenamefont {Krasheninnikov}}]{komsa2012two}%
  \BibitemOpen
  \bibfield  {author} {\bibinfo {author} {\bibfnamefont {H.-P.}\ \bibnamefont
  {Komsa}}, \bibinfo {author} {\bibfnamefont {J.}~\bibnamefont {Kotakoski}},
  \bibinfo {author} {\bibfnamefont {S.}~\bibnamefont {Kurasch}}, \bibinfo
  {author} {\bibfnamefont {O.}~\bibnamefont {Lehtinen}}, \bibinfo {author}
  {\bibfnamefont {U.}~\bibnamefont {Kaiser}}, \ and\ \bibinfo {author}
  {\bibfnamefont {A.~V.}\ \bibnamefont {Krasheninnikov}},\ }\href@noop {}
  {\bibfield  {journal} {\bibinfo  {journal} {Phys. Rev. Lett.}\ }\textbf
  {\bibinfo {volume} {109}},\ \bibinfo {pages} {035503} (\bibinfo {year}
  {2012})}\BibitemShut {NoStop}%
\bibitem [{\citenamefont {Ganatra}\ and\ \citenamefont
  {Zhang}(2014)}]{ganatra2014few}%
  \BibitemOpen
  \bibfield  {author} {\bibinfo {author} {\bibfnamefont {R.}~\bibnamefont
  {Ganatra}}\ and\ \bibinfo {author} {\bibfnamefont {Q.}~\bibnamefont
  {Zhang}},\ }\href@noop {} {\bibfield  {journal} {\bibinfo  {journal} {ACS
  Nano}\ }\textbf {\bibinfo {volume} {8}},\ \bibinfo {pages} {4074} (\bibinfo
  {year} {2014})}\BibitemShut {NoStop}%
\bibitem [{\citenamefont {Tsai}\ \emph {et~al.}(2014)\citenamefont {Tsai},
  \citenamefont {Su}, \citenamefont {Chang}, \citenamefont {Tsai},
  \citenamefont {Chen}, \citenamefont {Wu}, \citenamefont {Li}, \citenamefont
  {Chen},\ and\ \citenamefont {He}}]{tsai2014monolayer}%
  \BibitemOpen
  \bibfield  {author} {\bibinfo {author} {\bibfnamefont {M.-L.}\ \bibnamefont
  {Tsai}}, \bibinfo {author} {\bibfnamefont {S.-H.}\ \bibnamefont {Su}},
  \bibinfo {author} {\bibfnamefont {J.-K.}\ \bibnamefont {Chang}}, \bibinfo
  {author} {\bibfnamefont {D.-S.}\ \bibnamefont {Tsai}}, \bibinfo {author}
  {\bibfnamefont {C.-H.}\ \bibnamefont {Chen}}, \bibinfo {author}
  {\bibfnamefont {C.-I.}\ \bibnamefont {Wu}}, \bibinfo {author} {\bibfnamefont
  {L.-J.}\ \bibnamefont {Li}}, \bibinfo {author} {\bibfnamefont {L.-J.}\
  \bibnamefont {Chen}}, \ and\ \bibinfo {author} {\bibfnamefont {J.-H.}\
  \bibnamefont {He}},\ }\href@noop {} {\bibfield  {journal} {\bibinfo
  {journal} {ACS Nano}\ }\textbf {\bibinfo {volume} {8}},\ \bibinfo {pages}
  {8317} (\bibinfo {year} {2014})}\BibitemShut {NoStop}%
\bibitem [{\citenamefont {Gu}\ \emph {et~al.}(2013)\citenamefont {Gu},
  \citenamefont {Cui}, \citenamefont {Li}, \citenamefont {Wu}, \citenamefont
  {Zeng}, \citenamefont {Lee}, \citenamefont {Zhang},\ and\ \citenamefont
  {Sun}}]{gu2013solution}%
  \BibitemOpen
  \bibfield  {author} {\bibinfo {author} {\bibfnamefont {X.}~\bibnamefont
  {Gu}}, \bibinfo {author} {\bibfnamefont {W.}~\bibnamefont {Cui}}, \bibinfo
  {author} {\bibfnamefont {H.}~\bibnamefont {Li}}, \bibinfo {author}
  {\bibfnamefont {Z.}~\bibnamefont {Wu}}, \bibinfo {author} {\bibfnamefont
  {Z.}~\bibnamefont {Zeng}}, \bibinfo {author} {\bibfnamefont {S.-T.}\
  \bibnamefont {Lee}}, \bibinfo {author} {\bibfnamefont {H.}~\bibnamefont
  {Zhang}}, \ and\ \bibinfo {author} {\bibfnamefont {B.}~\bibnamefont {Sun}},\
  }\href@noop {} {\bibfield  {journal} {\bibinfo  {journal} {Adv. Energy
  Mater.}\ }\textbf {\bibinfo {volume} {3}},\ \bibinfo {pages} {1262} (\bibinfo
  {year} {2013})}\BibitemShut {NoStop}%
\bibitem [{\citenamefont {Bernardi}\ \emph {et~al.}(2013)\citenamefont
  {Bernardi}, \citenamefont {Palummo},\ and\ \citenamefont
  {Grossman}}]{bernardi2013extraordinary}%
  \BibitemOpen
  \bibfield  {author} {\bibinfo {author} {\bibfnamefont {M.}~\bibnamefont
  {Bernardi}}, \bibinfo {author} {\bibfnamefont {M.}~\bibnamefont {Palummo}}, \
  and\ \bibinfo {author} {\bibfnamefont {J.~C.}\ \bibnamefont {Grossman}},\
  }\href@noop {} {\bibfield  {journal} {\bibinfo  {journal} {Nano Lett.}\
  }\textbf {\bibinfo {volume} {13}},\ \bibinfo {pages} {3664} (\bibinfo {year}
  {2013})}\BibitemShut {NoStop}%
\bibitem [{\citenamefont {Karlick{\`y}}\ and\ \citenamefont
  {Otyepka}(2013)}]{karlicky2013band}%
  \BibitemOpen
  \bibfield  {author} {\bibinfo {author} {\bibfnamefont {F.}~\bibnamefont
  {Karlick{\`y}}}\ and\ \bibinfo {author} {\bibfnamefont {M.}~\bibnamefont
  {Otyepka}},\ }\href@noop {} {\bibfield  {journal} {\bibinfo  {journal} {J.
  Chem. Theory Comput.}\ }\textbf {\bibinfo {volume} {9}},\ \bibinfo {pages}
  {4155} (\bibinfo {year} {2013})}\BibitemShut {NoStop}%
\bibitem [{\citenamefont {Choi}\ \emph {et~al.}(2015)\citenamefont {Choi},
  \citenamefont {Cui}, \citenamefont {Lan},\ and\ \citenamefont
  {Zhang}}]{choi2015linear}%
  \BibitemOpen
  \bibfield  {author} {\bibinfo {author} {\bibfnamefont {J.-H.}\ \bibnamefont
  {Choi}}, \bibinfo {author} {\bibfnamefont {P.}~\bibnamefont {Cui}}, \bibinfo
  {author} {\bibfnamefont {H.}~\bibnamefont {Lan}}, \ and\ \bibinfo {author}
  {\bibfnamefont {Z.}~\bibnamefont {Zhang}},\ }\href@noop {} {\bibfield
  {journal} {\bibinfo  {journal} {Phys. Rev. Lett.}\ }\textbf {\bibinfo
  {volume} {115}},\ \bibinfo {pages} {066403} (\bibinfo {year}
  {2015})}\BibitemShut {NoStop}%
\bibitem [{\citenamefont {Latini}\ \emph {et~al.}(2015)\citenamefont {Latini},
  \citenamefont {Olsen},\ and\ \citenamefont {Thygesen}}]{latini2015excitons}%
  \BibitemOpen
  \bibfield  {author} {\bibinfo {author} {\bibfnamefont {S.}~\bibnamefont
  {Latini}}, \bibinfo {author} {\bibfnamefont {T.}~\bibnamefont {Olsen}}, \
  and\ \bibinfo {author} {\bibfnamefont {K.~S.}\ \bibnamefont {Thygesen}},\
  }\href@noop {} {\bibfield  {journal} {\bibinfo  {journal} {Phys. Rev. B}\
  }\textbf {\bibinfo {volume} {92}},\ \bibinfo {pages} {245123} (\bibinfo
  {year} {2015})}\BibitemShut {NoStop}%
\bibitem [{\citenamefont {Hedin}(1965)}]{hedin1965new}%
  \BibitemOpen
  \bibfield  {author} {\bibinfo {author} {\bibfnamefont {L.}~\bibnamefont
  {Hedin}},\ }\href@noop {} {\bibfield  {journal} {\bibinfo  {journal} {Phys.
  Rev.}\ }\textbf {\bibinfo {volume} {139}},\ \bibinfo {pages} {A796} (\bibinfo
  {year} {1965})}\BibitemShut {NoStop}%
\bibitem [{\citenamefont {Salpeter}\ and\ \citenamefont
  {Bethe}(1951)}]{salpeter1951relativistic}%
  \BibitemOpen
  \bibfield  {author} {\bibinfo {author} {\bibfnamefont {E.~E.}\ \bibnamefont
  {Salpeter}}\ and\ \bibinfo {author} {\bibfnamefont {H.~A.}\ \bibnamefont
  {Bethe}},\ }\href@noop {} {\bibfield  {journal} {\bibinfo  {journal} {Phys.
  Rev.}\ }\textbf {\bibinfo {volume} {84}},\ \bibinfo {pages} {1232} (\bibinfo
  {year} {1951})}\BibitemShut {NoStop}%
\bibitem [{\citenamefont {Karlick{\`y}}\ \emph {et~al.}(2012)\citenamefont
  {Karlick{\`y}}, \citenamefont {Zbo{\v{r}}il},\ and\ \citenamefont
  {Otyepka}}]{karlicky2012band}%
  \BibitemOpen
  \bibfield  {author} {\bibinfo {author} {\bibfnamefont {F.}~\bibnamefont
  {Karlick{\`y}}}, \bibinfo {author} {\bibfnamefont {R.}~\bibnamefont
  {Zbo{\v{r}}il}}, \ and\ \bibinfo {author} {\bibfnamefont {M.}~\bibnamefont
  {Otyepka}},\ }\href@noop {} {\bibfield  {journal} {\bibinfo  {journal} {J.
  Chem. Phys}\ }\textbf {\bibinfo {volume} {137}},\ \bibinfo {pages} {034709}
  (\bibinfo {year} {2012})}\BibitemShut {NoStop}%
\bibitem [{\citenamefont {Dovesi}\ \emph
  {et~al.}(2014{\natexlab{a}})\citenamefont {Dovesi}, \citenamefont {Orlando},
  \citenamefont {Erba}, \citenamefont {Zicovich-Wilson}, \citenamefont
  {Civalleri}, \citenamefont {Casassa}, \citenamefont {Maschio}, \citenamefont
  {Ferrabone}, \citenamefont {De~La~Pierre}, \citenamefont {D'Arco},
  \citenamefont {No\"{e}l}, \citenamefont {Caus\`{a}}, \citenamefont {Rerat},\
  and\ \citenamefont {Kirtman}}]{dovesi2014crystal14}%
  \BibitemOpen
  \bibfield  {author} {\bibinfo {author} {\bibfnamefont {R.}~\bibnamefont
  {Dovesi}}, \bibinfo {author} {\bibfnamefont {R.}~\bibnamefont {Orlando}},
  \bibinfo {author} {\bibfnamefont {A.}~\bibnamefont {Erba}}, \bibinfo {author}
  {\bibfnamefont {C.~M.}\ \bibnamefont {Zicovich-Wilson}}, \bibinfo {author}
  {\bibfnamefont {B.}~\bibnamefont {Civalleri}}, \bibinfo {author}
  {\bibfnamefont {S.}~\bibnamefont {Casassa}}, \bibinfo {author} {\bibfnamefont
  {L.}~\bibnamefont {Maschio}}, \bibinfo {author} {\bibfnamefont
  {M.}~\bibnamefont {Ferrabone}}, \bibinfo {author} {\bibfnamefont
  {M.}~\bibnamefont {De~La~Pierre}}, \bibinfo {author} {\bibfnamefont
  {P.}~\bibnamefont {D'Arco}}, \bibinfo {author} {\bibfnamefont
  {Y.}~\bibnamefont {No\"{e}l}}, \bibinfo {author} {\bibfnamefont
  {M.}~\bibnamefont {Caus\`{a}}}, \bibinfo {author} {\bibfnamefont
  {M.}~\bibnamefont {Rerat}}, \ and\ \bibinfo {author} {\bibfnamefont
  {B.}~\bibnamefont {Kirtman}},\ }\href@noop {} {\bibfield  {journal} {\bibinfo
   {journal} {Int. J. Quant. Chem.}\ }\textbf {\bibinfo {volume} {114}},\
  \bibinfo {pages} {1287} (\bibinfo {year} {2014}{\natexlab{a}})}\BibitemShut
  {NoStop}%
\bibitem [{\citenamefont {Dovesi}\ \emph
  {et~al.}(2014{\natexlab{b}})\citenamefont {Dovesi}, \citenamefont {Saunders},
  \citenamefont {Roetti}, \citenamefont {Orlando}, \citenamefont
  {Zicovich-Wilson}, \citenamefont {Pascale}, \citenamefont {Civalleri},
  \citenamefont {Doll}, \citenamefont {Harrison}, \citenamefont {Bush},
  \citenamefont {D'Arco}, \citenamefont {Llunell}, \citenamefont {Caus\`{a}},\
  and\ \citenamefont {No\"{e}l}}]{crystal14man}%
  \BibitemOpen
  \bibfield  {author} {\bibinfo {author} {\bibfnamefont {R.}~\bibnamefont
  {Dovesi}}, \bibinfo {author} {\bibfnamefont {V.~R.}\ \bibnamefont
  {Saunders}}, \bibinfo {author} {\bibfnamefont {C.}~\bibnamefont {Roetti}},
  \bibinfo {author} {\bibfnamefont {R.}~\bibnamefont {Orlando}}, \bibinfo
  {author} {\bibfnamefont {C.~M.}\ \bibnamefont {Zicovich-Wilson}}, \bibinfo
  {author} {\bibfnamefont {F.}~\bibnamefont {Pascale}}, \bibinfo {author}
  {\bibfnamefont {B.}~\bibnamefont {Civalleri}}, \bibinfo {author}
  {\bibfnamefont {K.}~\bibnamefont {Doll}}, \bibinfo {author} {\bibfnamefont
  {N.~M.}\ \bibnamefont {Harrison}}, \bibinfo {author} {\bibfnamefont {I.~J.}\
  \bibnamefont {Bush}}, \bibinfo {author} {\bibfnamefont {P.}~\bibnamefont
  {D'Arco}}, \bibinfo {author} {\bibfnamefont {M.}~\bibnamefont {Llunell}},
  \bibinfo {author} {\bibfnamefont {M.}~\bibnamefont {Caus\`{a}}}, \ and\
  \bibinfo {author} {\bibfnamefont {Y.}~\bibnamefont {No\"{e}l}},\ }\href@noop
  {} {\bibfield  {journal} {\bibinfo  {journal} {University of Torino: Torino}\
  } (\bibinfo {year} {2014}{\natexlab{b}})}\BibitemShut {NoStop}%
\bibitem [{\citenamefont {Zhao}\ and\ \citenamefont
  {Truhlar}(2008)}]{zhao2008m06}%
  \BibitemOpen
  \bibfield  {author} {\bibinfo {author} {\bibfnamefont {Y.}~\bibnamefont
  {Zhao}}\ and\ \bibinfo {author} {\bibfnamefont {D.~G.}\ \bibnamefont
  {Truhlar}},\ }\href@noop {} {\bibfield  {journal} {\bibinfo  {journal}
  {Theor. Chem. Acc.}\ }\textbf {\bibinfo {volume} {120}},\ \bibinfo {pages}
  {215} (\bibinfo {year} {2008})}\BibitemShut {NoStop}%
\bibitem [{\citenamefont {Peintinger}\ \emph {et~al.}(2013)\citenamefont
  {Peintinger}, \citenamefont {Oliveira},\ and\ \citenamefont
  {Bredow}}]{peintinger2013consistent}%
  \BibitemOpen
  \bibfield  {author} {\bibinfo {author} {\bibfnamefont {M.~F.}\ \bibnamefont
  {Peintinger}}, \bibinfo {author} {\bibfnamefont {D.~V.}\ \bibnamefont
  {Oliveira}}, \ and\ \bibinfo {author} {\bibfnamefont {T.}~\bibnamefont
  {Bredow}},\ }\href@noop {} {\bibfield  {journal} {\bibinfo  {journal} {J.
  Comput. Chem.}\ }\textbf {\bibinfo {volume} {34}},\ \bibinfo {pages} {451}
  (\bibinfo {year} {2013})}\BibitemShut {NoStop}%
\bibitem [{\citenamefont {Krukau}\ \emph {et~al.}(2006)\citenamefont {Krukau},
  \citenamefont {Vydrov}, \citenamefont {Izmaylov},\ and\ \citenamefont
  {Scuseria}}]{krukau2006influence}%
  \BibitemOpen
  \bibfield  {author} {\bibinfo {author} {\bibfnamefont {A.~V.}\ \bibnamefont
  {Krukau}}, \bibinfo {author} {\bibfnamefont {O.~A.}\ \bibnamefont {Vydrov}},
  \bibinfo {author} {\bibfnamefont {A.~F.}\ \bibnamefont {Izmaylov}}, \ and\
  \bibinfo {author} {\bibfnamefont {G.~E.}\ \bibnamefont {Scuseria}},\
  }\href@noop {} {\bibfield  {journal} {\bibinfo  {journal} {J. Chem. Phys.}\
  }\textbf {\bibinfo {volume} {125}},\ \bibinfo {pages} {224106} (\bibinfo
  {year} {2006})}\BibitemShut {NoStop}%
\bibitem [{\citenamefont {Heyd}\ \emph {et~al.}(2006)\citenamefont {Heyd},
  \citenamefont {Scuseria},\ and\ \citenamefont {Ernzerhof}}]{heyd2006erratum}%
  \BibitemOpen
  \bibfield  {author} {\bibinfo {author} {\bibfnamefont {J.}~\bibnamefont
  {Heyd}}, \bibinfo {author} {\bibfnamefont {G.~E.}\ \bibnamefont {Scuseria}},
  \ and\ \bibinfo {author} {\bibfnamefont {M.}~\bibnamefont {Ernzerhof}},\
  }\href@noop {} {\bibfield  {journal} {\bibinfo  {journal} {J. Chem. Phys.}\
  }\textbf {\bibinfo {volume} {124}},\ \bibinfo {pages} {219906} (\bibinfo
  {year} {2006})}\BibitemShut {NoStop}%
\bibitem [{\citenamefont {Heyd}\ and\ \citenamefont
  {Scuseria}(2004)}]{heyd2004efficient}%
  \BibitemOpen
  \bibfield  {author} {\bibinfo {author} {\bibfnamefont {J.}~\bibnamefont
  {Heyd}}\ and\ \bibinfo {author} {\bibfnamefont {G.~E.}\ \bibnamefont
  {Scuseria}},\ }\href@noop {} {\bibfield  {journal} {\bibinfo  {journal} {J.
  Chem. Phys.}\ }\textbf {\bibinfo {volume} {121}},\ \bibinfo {pages} {1187}
  (\bibinfo {year} {2004})}\BibitemShut {NoStop}%
\bibitem [{\citenamefont {Heyd}\ \emph {et~al.}(2003)\citenamefont {Heyd},
  \citenamefont {Scuseria},\ and\ \citenamefont {Ernzerhof}}]{heyd2003hybrid}%
  \BibitemOpen
  \bibfield  {author} {\bibinfo {author} {\bibfnamefont {J.}~\bibnamefont
  {Heyd}}, \bibinfo {author} {\bibfnamefont {G.~E.}\ \bibnamefont {Scuseria}},
  \ and\ \bibinfo {author} {\bibfnamefont {M.}~\bibnamefont {Ernzerhof}},\
  }\href@noop {} {\bibfield  {journal} {\bibinfo  {journal} {J. Chem. Phys.}\
  }\textbf {\bibinfo {volume} {118}},\ \bibinfo {pages} {8207} (\bibinfo {year}
  {2003})}\BibitemShut {NoStop}%
\bibitem [{\citenamefont {Steenbergen}\ \emph {et~al.}(2014)\citenamefont
  {Steenbergen}, \citenamefont {Gaston}, \citenamefont {M{\"u}ller},\ and\
  \citenamefont {Paulus}}]{steenbergen2014method}%
  \BibitemOpen
  \bibfield  {author} {\bibinfo {author} {\bibfnamefont {K.~G.}\ \bibnamefont
  {Steenbergen}}, \bibinfo {author} {\bibfnamefont {N.}~\bibnamefont {Gaston}},
  \bibinfo {author} {\bibfnamefont {C.}~\bibnamefont {M{\"u}ller}}, \ and\
  \bibinfo {author} {\bibfnamefont {B.}~\bibnamefont {Paulus}},\ }\href@noop {}
  {\bibfield  {journal} {\bibinfo  {journal} {J. Chem. Phys.}\ }\textbf
  {\bibinfo {volume} {141}},\ \bibinfo {pages} {124707} (\bibinfo {year}
  {2014})}\BibitemShut {NoStop}%
\bibitem [{\citenamefont {Dolg}\ \emph {et~al.}(1987)\citenamefont {Dolg},
  \citenamefont {Wedig}, \citenamefont {Stoll},\ and\ \citenamefont
  {Preuss}}]{dolg1987energy}%
  \BibitemOpen
  \bibfield  {author} {\bibinfo {author} {\bibfnamefont {M.}~\bibnamefont
  {Dolg}}, \bibinfo {author} {\bibfnamefont {U.}~\bibnamefont {Wedig}},
  \bibinfo {author} {\bibfnamefont {H.}~\bibnamefont {Stoll}}, \ and\ \bibinfo
  {author} {\bibfnamefont {H.}~\bibnamefont {Preuss}},\ }\href@noop {}
  {\bibfield  {journal} {\bibinfo  {journal} {J. Chem. Phys.}\ }\textbf
  {\bibinfo {volume} {86}},\ \bibinfo {pages} {866} (\bibinfo {year}
  {1987})}\BibitemShut {NoStop}%
\bibitem [{\citenamefont {Martin}\ and\ \citenamefont
  {Sundermann}(2001)}]{martin2001correlation}%
  \BibitemOpen
  \bibfield  {author} {\bibinfo {author} {\bibfnamefont {J.~M.}\ \bibnamefont
  {Martin}}\ and\ \bibinfo {author} {\bibfnamefont {A.}~\bibnamefont
  {Sundermann}},\ }\href@noop {} {\bibfield  {journal} {\bibinfo  {journal} {J.
  Chem. Phys.}\ }\textbf {\bibinfo {volume} {114}},\ \bibinfo {pages} {3408}
  (\bibinfo {year} {2001})}\BibitemShut {NoStop}%
\bibitem [{\citenamefont {Usvyat}(2015)}]{usvyat2015high}%
  \BibitemOpen
  \bibfield  {author} {\bibinfo {author} {\bibfnamefont {D.}~\bibnamefont
  {Usvyat}},\ }\href@noop {} {\bibfield  {journal} {\bibinfo  {journal} {J.
  Chem. Phys.}\ }\textbf {\bibinfo {volume} {143}},\ \bibinfo {pages} {104704}
  (\bibinfo {year} {2015})}\BibitemShut {NoStop}%
\bibitem [{\citenamefont {Yan}\ \emph {et~al.}(2012)\citenamefont {Yan},
  \citenamefont {Jacobsen},\ and\ \citenamefont {Thygesen}}]{yan2012optical}%
  \BibitemOpen
  \bibfield  {author} {\bibinfo {author} {\bibfnamefont {J.}~\bibnamefont
  {Yan}}, \bibinfo {author} {\bibfnamefont {K.~W.}\ \bibnamefont {Jacobsen}}, \
  and\ \bibinfo {author} {\bibfnamefont {K.~S.}\ \bibnamefont {Thygesen}},\
  }\href@noop {} {\bibfield  {journal} {\bibinfo  {journal} {Phys. Rev. B}\
  }\textbf {\bibinfo {volume} {86}},\ \bibinfo {pages} {045208} (\bibinfo
  {year} {2012})}\BibitemShut {NoStop}%
\bibitem [{\citenamefont {Olsen}\ \emph {et~al.}(2016)\citenamefont {Olsen},
  \citenamefont {Latini}, \citenamefont {Rasmussen},\ and\ \citenamefont
  {Thygesen}}]{olsen2016simple}%
  \BibitemOpen
  \bibfield  {author} {\bibinfo {author} {\bibfnamefont {T.}~\bibnamefont
  {Olsen}}, \bibinfo {author} {\bibfnamefont {S.}~\bibnamefont {Latini}},
  \bibinfo {author} {\bibfnamefont {F.}~\bibnamefont {Rasmussen}}, \ and\
  \bibinfo {author} {\bibfnamefont {K.~S.}\ \bibnamefont {Thygesen}},\
  }\href@noop {} {\bibfield  {journal} {\bibinfo  {journal} {Phys. Rev. Lett.}\
  }\textbf {\bibinfo {volume} {116}},\ \bibinfo {pages} {056401} (\bibinfo
  {year} {2016})}\BibitemShut {NoStop}%
\bibitem [{\citenamefont {Bahn}\ and\ \citenamefont
  {Jacobsen}(2002)}]{bahn2002object}%
  \BibitemOpen
  \bibfield  {author} {\bibinfo {author} {\bibfnamefont {S.~R.}\ \bibnamefont
  {Bahn}}\ and\ \bibinfo {author} {\bibfnamefont {K.~W.}\ \bibnamefont
  {Jacobsen}},\ }\href@noop {} {\bibfield  {journal} {\bibinfo  {journal}
  {Comp. Sci. Eng.}\ }\textbf {\bibinfo {volume} {4}},\ \bibinfo {pages} {56}
  (\bibinfo {year} {2002})}\BibitemShut {NoStop}%
\bibitem [{\citenamefont {Mortensen}\ \emph {et~al.}(2005)\citenamefont
  {Mortensen}, \citenamefont {Hansen},\ and\ \citenamefont
  {Jacobsen}}]{mortensen2005real}%
  \BibitemOpen
  \bibfield  {author} {\bibinfo {author} {\bibfnamefont {J.~J.}\ \bibnamefont
  {Mortensen}}, \bibinfo {author} {\bibfnamefont {L.~B.}\ \bibnamefont
  {Hansen}}, \ and\ \bibinfo {author} {\bibfnamefont {K.~W.}\ \bibnamefont
  {Jacobsen}},\ }\href@noop {} {\bibfield  {journal} {\bibinfo  {journal}
  {Phys. Rev. B}\ }\textbf {\bibinfo {volume} {71}},\ \bibinfo {pages} {035109}
  (\bibinfo {year} {2005})}\BibitemShut {NoStop}%
\bibitem [{\citenamefont {Enkovaara}\ \emph {et~al.}(2010)\citenamefont
  {Enkovaara}, \citenamefont {Rostgaard}, \citenamefont {Mortensen},
  \citenamefont {Chen}, \citenamefont {Du{\l}ak}, \citenamefont {Ferrighi},
  \citenamefont {Gavnholt}, \citenamefont {Glinsvad}, \citenamefont {Haikola},
  \citenamefont {Hansen}, \citenamefont {Kristoffersen}, \citenamefont
  {Kuisma}, \citenamefont {Larsen}, \citenamefont {Lehtovaara}, \citenamefont
  {Ljungberg}, \citenamefont {Lopez-Acevedo}, \citenamefont {Moses},
  \citenamefont {Ojanen}, \citenamefont {Olsen}, \citenamefont {Petzold},
  \citenamefont {Romero}, \citenamefont {Stausholm-M{\o}ller}, \citenamefont
  {Strange}, \citenamefont {Tritsaris}, \citenamefont {Vanin}, \citenamefont
  {Walter}, \citenamefont {Hammer}, \citenamefont {H\"{a}kkinen}, \citenamefont
  {Madsen}, \citenamefont {Nieminen}, \citenamefont {N{\o}rskov}, \citenamefont
  {Puska}, \citenamefont {Rantala}, \citenamefont {Schi{\o}tz}, \citenamefont
  {Thygesen},\ and\ \citenamefont {Jacobsen}}]{enkovaara2010electronic}%
  \BibitemOpen
  \bibfield  {author} {\bibinfo {author} {\bibfnamefont {J.}~\bibnamefont
  {Enkovaara}}, \bibinfo {author} {\bibfnamefont {C.}~\bibnamefont
  {Rostgaard}}, \bibinfo {author} {\bibfnamefont {J.~J.}\ \bibnamefont
  {Mortensen}}, \bibinfo {author} {\bibfnamefont {J.}~\bibnamefont {Chen}},
  \bibinfo {author} {\bibfnamefont {M.}~\bibnamefont {Du{\l}ak}}, \bibinfo
  {author} {\bibfnamefont {L.}~\bibnamefont {Ferrighi}}, \bibinfo {author}
  {\bibfnamefont {J.}~\bibnamefont {Gavnholt}}, \bibinfo {author}
  {\bibfnamefont {C.}~\bibnamefont {Glinsvad}}, \bibinfo {author}
  {\bibfnamefont {V.}~\bibnamefont {Haikola}}, \bibinfo {author} {\bibfnamefont
  {H.~A.}\ \bibnamefont {Hansen}}, \bibinfo {author} {\bibfnamefont {H.~H.}\
  \bibnamefont {Kristoffersen}}, \bibinfo {author} {\bibfnamefont
  {M.}~\bibnamefont {Kuisma}}, \bibinfo {author} {\bibfnamefont {A.~H.}\
  \bibnamefont {Larsen}}, \bibinfo {author} {\bibfnamefont {L.}~\bibnamefont
  {Lehtovaara}}, \bibinfo {author} {\bibfnamefont {M.}~\bibnamefont
  {Ljungberg}}, \bibinfo {author} {\bibfnamefont {O.}~\bibnamefont
  {Lopez-Acevedo}}, \bibinfo {author} {\bibfnamefont {P.~G.}\ \bibnamefont
  {Moses}}, \bibinfo {author} {\bibfnamefont {J.}~\bibnamefont {Ojanen}},
  \bibinfo {author} {\bibfnamefont {T.}~\bibnamefont {Olsen}}, \bibinfo
  {author} {\bibfnamefont {V.}~\bibnamefont {Petzold}}, \bibinfo {author}
  {\bibfnamefont {N.~A.}\ \bibnamefont {Romero}}, \bibinfo {author}
  {\bibfnamefont {J.}~\bibnamefont {Stausholm-M{\o}ller}}, \bibinfo {author}
  {\bibfnamefont {M.}~\bibnamefont {Strange}}, \bibinfo {author} {\bibfnamefont
  {G.~A.}\ \bibnamefont {Tritsaris}}, \bibinfo {author} {\bibfnamefont
  {M.}~\bibnamefont {Vanin}}, \bibinfo {author} {\bibfnamefont
  {M.}~\bibnamefont {Walter}}, \bibinfo {author} {\bibfnamefont
  {B.}~\bibnamefont {Hammer}}, \bibinfo {author} {\bibfnamefont
  {H.}~\bibnamefont {H\"{a}kkinen}}, \bibinfo {author} {\bibfnamefont
  {G.~K.~H.}\ \bibnamefont {Madsen}}, \bibinfo {author} {\bibfnamefont {R.~M.}\
  \bibnamefont {Nieminen}}, \bibinfo {author} {\bibfnamefont {J.~K.}\
  \bibnamefont {N{\o}rskov}}, \bibinfo {author} {\bibfnamefont
  {M.}~\bibnamefont {Puska}}, \bibinfo {author} {\bibfnamefont {T.~T.}\
  \bibnamefont {Rantala}}, \bibinfo {author} {\bibfnamefont {J.}~\bibnamefont
  {Schi{\o}tz}}, \bibinfo {author} {\bibfnamefont {K.~S.}\ \bibnamefont
  {Thygesen}}, \ and\ \bibinfo {author} {\bibfnamefont {K.~W.}\ \bibnamefont
  {Jacobsen}},\ }\href@noop {} {\bibfield  {journal} {\bibinfo  {journal} {J.
  Phys. Condens. Matter}\ }\textbf {\bibinfo {volume} {22}},\ \bibinfo {pages}
  {253202} (\bibinfo {year} {2010})}\BibitemShut {NoStop}%
\bibitem [{\citenamefont {Yan}\ \emph {et~al.}(2011)\citenamefont {Yan},
  \citenamefont {Mortensen}, \citenamefont {Jacobsen},\ and\ \citenamefont
  {Thygesen}}]{yan2011linear}%
  \BibitemOpen
  \bibfield  {author} {\bibinfo {author} {\bibfnamefont {J.}~\bibnamefont
  {Yan}}, \bibinfo {author} {\bibfnamefont {J.~J.}\ \bibnamefont {Mortensen}},
  \bibinfo {author} {\bibfnamefont {K.~W.}\ \bibnamefont {Jacobsen}}, \ and\
  \bibinfo {author} {\bibfnamefont {K.~S.}\ \bibnamefont {Thygesen}},\
  }\href@noop {} {\bibfield  {journal} {\bibinfo  {journal} {Phys. Rev. B}\
  }\textbf {\bibinfo {volume} {83}},\ \bibinfo {pages} {245122} (\bibinfo
  {year} {2011})}\BibitemShut {NoStop}%
\bibitem [{\citenamefont {H{\"u}ser}\ \emph {et~al.}(2013)\citenamefont
  {H{\"u}ser}, \citenamefont {Olsen},\ and\ \citenamefont
  {Thygesen}}]{huser2013quasiparticle}%
  \BibitemOpen
  \bibfield  {author} {\bibinfo {author} {\bibfnamefont {F.}~\bibnamefont
  {H{\"u}ser}}, \bibinfo {author} {\bibfnamefont {T.}~\bibnamefont {Olsen}}, \
  and\ \bibinfo {author} {\bibfnamefont {K.~S.}\ \bibnamefont {Thygesen}},\
  }\href@noop {} {\bibfield  {journal} {\bibinfo  {journal} {Phys. Rev. B}\
  }\textbf {\bibinfo {volume} {87}},\ \bibinfo {pages} {235132} (\bibinfo
  {year} {2013})}\BibitemShut {NoStop}%
\bibitem [{\citenamefont {Rasmussen}\ \emph {et~al.}(2016)\citenamefont
  {Rasmussen}, \citenamefont {Schmidt}, \citenamefont {Winther},\ and\
  \citenamefont {Thygesen}}]{rasmussen2015efficient}%
  \BibitemOpen
  \bibfield  {author} {\bibinfo {author} {\bibfnamefont {F.~A.}\ \bibnamefont
  {Rasmussen}}, \bibinfo {author} {\bibfnamefont {P.~S.}\ \bibnamefont
  {Schmidt}}, \bibinfo {author} {\bibfnamefont {K.~T.}\ \bibnamefont
  {Winther}}, \ and\ \bibinfo {author} {\bibfnamefont {K.~S.}\ \bibnamefont
  {Thygesen}},\ }\href@noop {} {\bibfield  {journal} {\bibinfo  {journal}
  {arXiv preprint arXiv:1511.00129}\ } (\bibinfo {year} {2016})}\BibitemShut
  {NoStop}%
\bibitem [{\citenamefont {Perdew}\ \emph {et~al.}(1996)\citenamefont {Perdew},
  \citenamefont {Burke},\ and\ \citenamefont
  {Ernzerhof}}]{perdew1996generalized}%
  \BibitemOpen
  \bibfield  {author} {\bibinfo {author} {\bibfnamefont {J.~P.}\ \bibnamefont
  {Perdew}}, \bibinfo {author} {\bibfnamefont {K.}~\bibnamefont {Burke}}, \
  and\ \bibinfo {author} {\bibfnamefont {M.}~\bibnamefont {Ernzerhof}},\
  }\href@noop {} {\bibfield  {journal} {\bibinfo  {journal} {Phys. Rev. Lett.}\
  }\textbf {\bibinfo {volume} {77}},\ \bibinfo {pages} {3865} (\bibinfo {year}
  {1996})}\BibitemShut {NoStop}%
\bibitem [{\citenamefont {Fuchs}\ \emph {et~al.}(2007)\citenamefont {Fuchs},
  \citenamefont {Furthm{\"u}ller}, \citenamefont {Bechstedt}, \citenamefont
  {Shishkin},\ and\ \citenamefont {Kresse}}]{fuchs2007quasiparticle}%
  \BibitemOpen
  \bibfield  {author} {\bibinfo {author} {\bibfnamefont {F.}~\bibnamefont
  {Fuchs}}, \bibinfo {author} {\bibfnamefont {J.}~\bibnamefont
  {Furthm{\"u}ller}}, \bibinfo {author} {\bibfnamefont {F.}~\bibnamefont
  {Bechstedt}}, \bibinfo {author} {\bibfnamefont {M.}~\bibnamefont {Shishkin}},
  \ and\ \bibinfo {author} {\bibfnamefont {G.}~\bibnamefont {Kresse}},\
  }\href@noop {} {\bibfield  {journal} {\bibinfo  {journal} {Phys. Rev. B}\
  }\textbf {\bibinfo {volume} {76}},\ \bibinfo {pages} {115109} (\bibinfo
  {year} {2007})}\BibitemShut {NoStop}%
\bibitem [{\citenamefont {Kuisma}\ \emph {et~al.}(2010)\citenamefont {Kuisma},
  \citenamefont {Ojanen}, \citenamefont {Enkovaara},\ and\ \citenamefont
  {Rantala}}]{kuisma2010kohn}%
  \BibitemOpen
  \bibfield  {author} {\bibinfo {author} {\bibfnamefont {M.}~\bibnamefont
  {Kuisma}}, \bibinfo {author} {\bibfnamefont {J.}~\bibnamefont {Ojanen}},
  \bibinfo {author} {\bibfnamefont {J.}~\bibnamefont {Enkovaara}}, \ and\
  \bibinfo {author} {\bibfnamefont {T.~T.}\ \bibnamefont {Rantala}},\
  }\href@noop {} {\bibfield  {journal} {\bibinfo  {journal} {Phys. Rev. B}\
  }\textbf {\bibinfo {volume} {82}},\ \bibinfo {pages} {115106} (\bibinfo
  {year} {2010})}\BibitemShut {NoStop}%
\bibitem [{\citenamefont {Perdew}\ \emph {et~al.}(1982)\citenamefont {Perdew},
  \citenamefont {Parr}, \citenamefont {Levy},\ and\ \citenamefont
  {Balduz~Jr.}}]{perdew1982density}%
  \BibitemOpen
  \bibfield  {author} {\bibinfo {author} {\bibfnamefont {J.~P.}\ \bibnamefont
  {Perdew}}, \bibinfo {author} {\bibfnamefont {R.~G.}\ \bibnamefont {Parr}},
  \bibinfo {author} {\bibfnamefont {M.}~\bibnamefont {Levy}}, \ and\ \bibinfo
  {author} {\bibfnamefont {J.~L.}\ \bibnamefont {Balduz~Jr.}},\ }\href@noop {}
  {\bibfield  {journal} {\bibinfo  {journal} {Phys. Rev. Lett.}\ }\textbf
  {\bibinfo {volume} {49}},\ \bibinfo {pages} {1691} (\bibinfo {year}
  {1982})}\BibitemShut {NoStop}%
\bibitem [{\citenamefont {Castelli}\ \emph {et~al.}(2012)\citenamefont
  {Castelli}, \citenamefont {Olsen}, \citenamefont {Datta}, \citenamefont
  {Landis}, \citenamefont {Dahl}, \citenamefont {Thygesen},\ and\ \citenamefont
  {Jacobsen}}]{castelli2012computational}%
  \BibitemOpen
  \bibfield  {author} {\bibinfo {author} {\bibfnamefont {I.~E.}\ \bibnamefont
  {Castelli}}, \bibinfo {author} {\bibfnamefont {T.}~\bibnamefont {Olsen}},
  \bibinfo {author} {\bibfnamefont {S.}~\bibnamefont {Datta}}, \bibinfo
  {author} {\bibfnamefont {D.~D.}\ \bibnamefont {Landis}}, \bibinfo {author}
  {\bibfnamefont {S.}~\bibnamefont {Dahl}}, \bibinfo {author} {\bibfnamefont
  {K.~S.}\ \bibnamefont {Thygesen}}, \ and\ \bibinfo {author} {\bibfnamefont
  {K.~W.}\ \bibnamefont {Jacobsen}},\ }\href@noop {} {\bibfield  {journal}
  {\bibinfo  {journal} {Energy \& Environ. Sci.}\ }\textbf {\bibinfo {volume}
  {5}},\ \bibinfo {pages} {5814} (\bibinfo {year} {2012})}\BibitemShut
  {NoStop}%
\bibitem [{\citenamefont {Mori-S{\'a}nchez}\ \emph {et~al.}(2008)\citenamefont
  {Mori-S{\'a}nchez}, \citenamefont {Cohen},\ and\ \citenamefont
  {Yang}}]{mori2008localization}%
  \BibitemOpen
  \bibfield  {author} {\bibinfo {author} {\bibfnamefont {P.}~\bibnamefont
  {Mori-S{\'a}nchez}}, \bibinfo {author} {\bibfnamefont {A.~J.}\ \bibnamefont
  {Cohen}}, \ and\ \bibinfo {author} {\bibfnamefont {W.}~\bibnamefont {Yang}},\
  }\href@noop {} {\bibfield  {journal} {\bibinfo  {journal} {Phys. Rev. Lett.}\
  }\textbf {\bibinfo {volume} {100}},\ \bibinfo {pages} {146401} (\bibinfo
  {year} {2008})}\BibitemShut {NoStop}%
\bibitem [{\citenamefont {Shockley}\ and\ \citenamefont
  {Queisser}(1961)}]{shockley1961detailed}%
  \BibitemOpen
  \bibfield  {author} {\bibinfo {author} {\bibfnamefont {W.}~\bibnamefont
  {Shockley}}\ and\ \bibinfo {author} {\bibfnamefont {H.~J.}\ \bibnamefont
  {Queisser}},\ }\href@noop {} {\bibfield  {journal} {\bibinfo  {journal} {J.
  Appl. Phys.}\ }\textbf {\bibinfo {volume} {32}},\ \bibinfo {pages} {510}
  (\bibinfo {year} {1961})}\BibitemShut {NoStop}%
\bibitem [{\citenamefont {Miller}\ \emph {et~al.}(2012)\citenamefont {Miller},
  \citenamefont {Yablonovitch},\ and\ \citenamefont
  {Kurtz}}]{miller2012strong}%
  \BibitemOpen
  \bibfield  {author} {\bibinfo {author} {\bibfnamefont {O.~D.}\ \bibnamefont
  {Miller}}, \bibinfo {author} {\bibfnamefont {E.}~\bibnamefont
  {Yablonovitch}}, \ and\ \bibinfo {author} {\bibfnamefont {S.~R.}\
  \bibnamefont {Kurtz}},\ }\href@noop {} {\bibfield  {journal} {\bibinfo
  {journal} {IEEE J. Photovolt.}\ }\textbf {\bibinfo {volume} {2}},\ \bibinfo
  {pages} {303} (\bibinfo {year} {2012})}\BibitemShut {NoStop}%
\bibitem [{\citenamefont {Momma}\ and\ \citenamefont
  {Izumi}(2011)}]{momma2011vesta}%
  \BibitemOpen
  \bibfield  {author} {\bibinfo {author} {\bibfnamefont {K.}~\bibnamefont
  {Momma}}\ and\ \bibinfo {author} {\bibfnamefont {F.}~\bibnamefont {Izumi}},\
  }\href@noop {} {\bibfield  {journal} {\bibinfo  {journal} {J. Appl.
  Crystallogr.}\ }\textbf {\bibinfo {volume} {44}},\ \bibinfo {pages} {1272}
  (\bibinfo {year} {2011})}\BibitemShut {NoStop}%
\bibitem [{\citenamefont {Hunter}(2007)}]{Hunter_2007}%
  \BibitemOpen
  \bibfield  {author} {\bibinfo {author} {\bibfnamefont {J.~D.}\ \bibnamefont
  {Hunter}},\ }\href@noop {} {\bibfield  {journal} {\bibinfo  {journal}
  {Comput. Sci. Eng.}\ }\textbf {\bibinfo {volume} {9}},\ \bibinfo {pages} {90}
  (\bibinfo {year} {2007})}\BibitemShut {NoStop}%
\bibitem [{\citenamefont {Rozzi}\ \emph {et~al.}(2006)\citenamefont {Rozzi},
  \citenamefont {Varsano}, \citenamefont {Marini}, \citenamefont {Gross},\ and\
  \citenamefont {Rubio}}]{rozzi2006exact}%
  \BibitemOpen
  \bibfield  {author} {\bibinfo {author} {\bibfnamefont {C.~A.}\ \bibnamefont
  {Rozzi}}, \bibinfo {author} {\bibfnamefont {D.}~\bibnamefont {Varsano}},
  \bibinfo {author} {\bibfnamefont {A.}~\bibnamefont {Marini}}, \bibinfo
  {author} {\bibfnamefont {E.~K.~U.}\ \bibnamefont {Gross}}, \ and\ \bibinfo
  {author} {\bibfnamefont {A.}~\bibnamefont {Rubio}},\ }\href@noop {}
  {\bibfield  {journal} {\bibinfo  {journal} {Phys. Rev. B}\ }\textbf {\bibinfo
  {volume} {73}},\ \bibinfo {pages} {205119} (\bibinfo {year}
  {2006})}\BibitemShut {NoStop}%
\bibitem [{\citenamefont {Sundararaman}\ and\ \citenamefont
  {Arias}(2013)}]{sundararaman2013regularization}%
  \BibitemOpen
  \bibfield  {author} {\bibinfo {author} {\bibfnamefont {R.}~\bibnamefont
  {Sundararaman}}\ and\ \bibinfo {author} {\bibfnamefont {T.~A.}\ \bibnamefont
  {Arias}},\ }\href@noop {} {\bibfield  {journal} {\bibinfo  {journal} {Phys.
  Rev. B}\ }\textbf {\bibinfo {volume} {87}},\ \bibinfo {pages} {165122}
  (\bibinfo {year} {2013})}\BibitemShut {NoStop}%
\end{thebibliography}%
\newpage
\appendix
\section{Supporting information for Strong 1D localization and highly anisotropic electron-hole masses in heavy-halogen functionalized graphenes}
\label{supporting}

\subsection*{GPAW}

DFT band structure calculations were performed using a real-space grid description of the wave function with a grid spacing of $\approx 0.14$ \AA. The resulting band gaps were found to agree with plane-wave calculations using a 600 eV cut-off to within better then $0.01$ eV.

Ground-state calculations preceding $G_0W_0$ calculations were done using an energy cut-off of 600 eV while $G_0W_0$ self energies were calculated at three cut-off values up to 120 eV (103 eV, 111 eV, 120 eV) and extrapolated to infinite cut-off. $G_0W_0$ calculations employed a $10\times10$ k-point sampling. The frequency dependence was represented on a non-linear grid from 0 eV to the energy of the highest transition included in the basis set where the grid-spacing was gradually increased starting from 0.1 eV and reaching 0.2 eV at 4.0 eV. A vacuum of 10 \AA\ was used to separate periodic images of the slabs along the surface-normal directions. $G_0W_0$ calculations employed a 2D cut-off of the Coulomb interaction along the surface-normal direction including analytical corrections for the $q\rightarrow 0$ contributions\cite{rasmussen2015efficient,rozzi2006exact}. Exchange contributions were evaluated using Wigner-Seitz truncation\cite{sundararaman2013regularization}.

BSE calculations require significant care, especially when a truncated Coulomb interaction is used, as the screened potential converges very slowly with the k-space sampling. We found that in order to converge the first optical excitation to within $\approx 0.05$ eV it was necessary to use k-grids of $18\times18$. The convergence behavior with respect to the energy cut-off and number of bands included in calculating the screened interaction was found to be much more favorable though and a cut-off value of 40 eV was sufficient as increasing the cut-off value to 150 eV only changed the position of the first excitation peak by $\approx 0.02$ eV. The construction of the BSE-Hamiltonian included bands up to 5 eV from the valence band maximum and conduction band minimum respectively. All BSE calculations are performed using the statically screened interaction evaluated within the random phase approximation and employing the Tamm-Dancoff approximation. A vacuum of 12 \AA\ was used to separate periodic images of the slabs along the surface-normal directions. BSE calculations employed a 2D cut-off of the Coulomb interaction along the surface-normal direction\cite{rozzi2006exact}.

\subsection*{CRYSTAL}

All CRYSTAL calculations employed a $24\times24$ k-grid sampling while CRYSTAL thresholds (TOLINTEG) were set to $10^{-8}$, $10^{-8}$, $10^{-8}$ and $10^{-8}$, $10^{-16}$ respectively and SETINF values 41 and 43 were set to 30 and 20 respectively (for more information on these thresholds see the CRYSTAL manual on the website \url{www.crystal.unito.it}).

Structure optimizations on all systems were performed using the CRYSTAL14 program\cite{dovesi2014crystal14,crystal14man} together with the M06-2X\cite{zhao2008m06} functional using the \mbox{POB-triple-$\zeta$} basis set proposed by Peintinger et al.\cite{peintinger2013consistent}. In the case of Br and Cl, HSE03\cite{krukau2006influence, heyd2006erratum, heyd2004efficient, heyd2003hybrid} band gap calculations on the relaxed structures were done employing the Stuttgart triple-$\zeta$ basis set as modified for use in periodic calculations by Steenbergen et al.\cite{steenbergen2014method}, together with the associated quasirelativistic pseudopotentials\cite{dolg1987energy, martin2001correlation}. For C, F and H, basis sets were constructed according to the procedure described by Usvyat\cite{usvyat2015high}. In all cases the description of the vacuum region was enhanced by adding ghost atoms containing a 1s function with an exponent of 0.06 $a_0^ {-1}$, 1 \AA\ above the position of the halogen atoms. The complete basis set used is given below.\\ \\
\noindent
\textbf{2 1}\\\noindent
0 0 1 2.0 1.0\\
      0.06              1.0000000\\
\textbf{1 6}\\
0 0 3 1.0 1.0\\
     34.0613410              0.60251978E-02\\
      5.1235746              0.45021094E-01\\
      1.1646626              0.20189726\\
0 0 1 0.0 1.0\\
      0.32723041             1.0000000\\
0 0 1 0.0 1.0\\
      0.10307241             1.0000000\\
0 2 1 0.0 1.0\\
      1.4070000              1.0000000\\
0 2 1 0.0 1.0\\
      0.3880000              1.0000000\\
0 3 1 0.0 1.0\\
      1.0570000              1.0000000\\
\textbf{6 11}\\
0 0 5 2.0 1.0\\
   8506.0384000              0.53373664E-03   \\
   1275.7329000              0.41250232E-02  \\ 
    290.3118700              0.21171337E-01  \\ 
     82.0562000              0.82417860E-01  \\ 
     26.4796410              0.24012858      \\ 
0 0 1 2.0 1.0\\
      9.2414585              1.0000000       \\ 
0 0 1 0.0 1.0\\
      3.3643530              1.0000000      \\  
0 0 1 0.0 1.0\\
      0.87174164             1.0000000     \\   
0 0 1 0.0 1.0\\
      0.36352352             1.0000000  \\      
0 0 1 0.0 1.0\\
      0.12873135             1.0000000   \\     
0 2 4 2.0 1.0\\
     34.7094960              0.53300974E-02   \\
      7.9590883              0.35865814E-01  \\ 
      2.3786972              0.14200299    \\   
      0.81540065             0.34203105   \\    
0 2 1 0.0 1.0\\
      0.28953785             1.0000000    \\    
0 3 1 0.0 1.0\\
      1.0970000              1.0000000   \\     
0 3 1 0.0 1.0\\
      0.3180000              1.0000000   \\     
0 3 1 0.0 1.0\\
      0.7610000              1.0000000    \\ 
\textbf{9 12}\\
0 0 5 2.0 1.0\\
  20450.4890000              0.51103495E-03 \\  
   3066.9547000              0.39518820E-02 \\  
    697.9100300              0.20334553E-01 \\  
    197.2702000              0.79876480E-01  \\ 
     63.7283430              0.23775601   \\    
0 0 1 2.0 1.0\\
     22.3218090              1.0000000   \\     
0 0 1 0.0 1.0\\
      8.1557609              1.0000000   \\     
0 0 1 0.0 1.0\\
      2.2114295              1.0000000   \\     
0 0 1 0.0 1.0\\
      0.89038567             1.0000000   \\     
0 0 1 0.0 1.0\\
      0.30696604             1.0000000    \\    
0 2 4 5.0 1.0\\
     80.2180200              0.63744464E-02 \\  
     18.5872810              0.44360191E-01 \\  
      5.6844581              0.16880038   \\    
      1.9512781              0.36162979  \\     
0 2 1 0.0 1.0\\
      0.67024114             1.0000000   \\     
0 2 1 0.0 1.0\\
      0.21682252             1.0000000  \\     
0 3 1 0.0 1.0\\
      3.1070000              1.0000000 \\       
0 3 1 0.0 1.0\\
      0.8550000              1.0000000  \\      
0 4 1 0.0 1.0\\
      1.9170000              1.0000000 \\   
\textbf{217 12}\\
INPUT\\
7. 0 2 2 1 0 0\\
6.394300 33.136632 0\\
3.197100 16.270728 0\\
5.620700 24.416993 0\\
2.810300 7.683050 0\\
5.338100 -8.587649 0\\
0 0 3 2.0 1.0\\
104.3829980 0.0031560\\
10.9005580 0.0239720\\
2.2685170 -0.3310080\\
0 0 1 0.0 1.0\\
0.9567350 1.0000000\\
0 0 1 0.0 1.0\\
0.3943800 1.0000000\\
0 0 1 0.0 1.0\\
0.1380120 1.0000000\\
0 2 3 5.0 1.0\\
17.9293820 0.0029790\\
3.2048610 -0.0600800\\
1.5221960 0.0690590\\
0 2 1 0.0 1.0\\
0.6753990 1.0000000\\
0 2 1 0.0 1.0\\
0.2541180 1.0000000\\
0 3 1 0.0 1.0\\
1.0460000 1.0000000\\
0 3 1 0.0 1.0\\
0.5440000 1.0000000\\
0 3 1 0.0 1.0\\
0.1350000 1.0000000\\
0 4 1 0.0 1.0\\
0.7060000 1.0000000\\
0 4 1 0.0 1.0\\
0.3120000 1.0000000\\
\textbf{235 11}\\
INPUT\\
7. 0 2 2 2 1 0\\
5.021800 61.513721 0\\
2.510900 9.021493 0\\
4.281400 53.875864 0\\
2.140700 4.629402 0\\
2.880000 20.849677 0\\
1.440000 2.965444 0\\
2.720700 -8.161493 0\\
0 0 10 2. 1.\\
762.0066790 0.0000520\\
376.6365900 -0.0000230\\
186.1599440 0.0001490\\
92.0131660 0.0001420\\
45.4792930 0.0001820\\
22.4790230 0.0014490\\
11.1106930 -0.0118330\\
5.4916760 0.1119000\\
2.7143670 -0.2561560\\
1.3416290 -0.4007910\\
0 0 1 0.0 1.0\\
0.9528400 1.0000000\\
0 0 1 0.0 1.0\\
0.3988620 1.0000000\\
0 0 1 0.0 1.0\\
0.1919290 1.0000000\\
0 2 10 5.0 1.0\\
0.0641740 0.0906150\\
0.1442090 0.3433450\\
0.3131070 0.4482900\\
0.6656170 0.2831560\\
77.0646890 0.0000120\\
33.1900760 0.0001780\\
14.2942400 -0.0008620\\
6.1562170 0.0165350\\
2.6513480 -0.1173210\\
1.1418780 -0.0129850\\
0 2 1 0.0 1.0\\
0.6656170 1.0000000\\
0 2 1 0.0 1.0\\
0.1442090 1.0000000\\
0 3 1 0.0 1.0\\
0.6013860 1.0000000\\
0 3 1 0.0 1.0\\
0.2523660 1.0000000\\
0 4 1 0.0 1.0\\
0.5812580 1.0000000\\
0 4 1 0.0 1.0\\
0.2592890 1.0000000\\\\

\subsection*{Structural and stability data}

\begin{table}[h!]
\begin{center}
\begin{tabular}{@{}lcccc@{}} 
\hline
      & \multicolumn{2}{c}{M06-2X}  & \multicolumn{2}{c}{PBE} \\
        \cline{2-3}                \cline{4-5}                  

      & a($X$)& b($Y$)& a($X$) & b($Y$)  \\ \hline

BrHH  &  5.14 &  5.29 &  5.16 &  5.29  \\ 

BrHBr &  5.21 &  5.50 &  5.22 &  5.45\\  

BrHCl &  5.18 &  5.46 &  5.19 &  5.43\\ 

BrFF  &  5.29 &  5.49 &  5.19 &  5.18\\ 

BrFBr &  5.39 &  5.73 &  5.45 &  5.76\\  

BrFCl &  5.36 &  5.63 &  5.42 &  5.67\\ 

\hline
\end{tabular}
 \caption{M06-2X- and PBE-relaxed lattice constants along the $X$ and $Y$ direction (see figure \ref{m062x_structures} for definitions). All values are given in  \AA.}
\end{center}
\end{table}

\begin{table}[h!]
\begin{center}
\begin{tabular}{@{}lcccccccc@{}} 
\hline
       & \multicolumn{2}{c}{GrH} & \multicolumn{2}{c}{GrF} & \multicolumn{2}{c}{GrCl} & \multicolumn{2}{c}{GrBr}\\
\cline{2-3}    \cline{4-5}    \cline{6-7}    \cline{8-9} 
       &   M06-2X &    PBE &    M06-2X &    PBE &    M06-2X &    PBE &    M06-2X &    PBE\\ \hline

BrFF   &  -67 & -59 & 105 &  85 & -203 &    -205 &  -311 &  -303 \\

BrFBr  &   18 &  33 & 189 & 176 & -119 &    -113 & -227 &   -211 \\

BrFCl  &   -4 &  12 & 167 & 155 & -141 &    -134 &  -249 &  -232 \\

BrHH   &   43 &  40 & 214 & 184 &  -93 &    -106 &  -202 &  -204 \\

BrHBr  &   69 &  67 & 240 & 211 &  -68 &    -79  & -176 &   -177 \\

BrHCl  &   49 &  49 & 220 & 192 &  -88 &    -97  &  -196 &  -195 \\
\hline
\end{tabular}
 \caption{Stability of the compounds considered in this work as compared to GrH, GrF, GrCl and GrBr calculated using M06-2X and PBE. All values in kJ/mol normalized to the number of carbon atoms in the unit cell. For comparison the M06-2X stabilities of GrF, GrCl and GrBr relative to GrH are -172 kJ/mol, 136 kJ/mol and 245 kJ/mol respectively and -143 kJ/mol, 146 kJ/mol and 244 kJ/mol using PBE.}
\end{center}
\end{table}

\end{document}